\documentclass[%
superscriptaddress,
 amSIath,amssymb,
pre,
twocolumn,
]{revtex4-2}

\usepackage{xcolor}
\usepackage{graphicx}
\usepackage{dcolumn}
\usepackage{bm}
\usepackage[normalem]{ulem} 

\usepackage{amsmath}
\usepackage{amsfonts}
\usepackage{amssymb}
\usepackage{epsfig}
\usepackage{graphicx}

\renewcommand{\phi}{\varphi}

\newcommand{\be}{\begin{equation}}
\newcommand{\ee}{\end{equation}}
\newcommand{\bea}{\begin{equnaray}}
\newcommand{\eea}{\end{equnaray}}
\newcommand{\ba}{\begin{align}}
\newcommand{\ea}{\end{align}}

\usepackage{color}

\newcommand{\beginsupplement}{
	\onecolumngrid
        \clearpage
        \setcounter{table}{0}
        \renewcommand{\thetable}{S\arabic{table}}
        \setcounter{figure}{0}
        \renewcommand{\thefigure}{S\arabic{figure}}
        \setcounter{section}{0}
        \renewcommand{\thesection}{S\arabic{section}}
     }

\begin{document}


\title{Creating bulk ultrastable glasses by random particle bonding}

\author{Misaki Ozawa}
\affiliation{Laboratoire de Physique de l'Ecole normale sup\'erieure, ENS, Universit\'e PSL, CNRS, Sorbonne Universit\'e, Universit\'e Paris-Diderot, Sorbonne Paris Cit\'e, Paris, France}

\author{Yasutaka Iwashita}
\affiliation{Department of Physics, Kyoto Sangyo University, Kyoto, Japan}

\author{Walter Kob}
\affiliation{Laboratoire Charles Coulomb, University of Montpellier and CNRS, F-34095 Montpellier,France}

\author{Francesco Zamponi}
\affiliation{Laboratoire de Physique de l'Ecole normale sup\'erieure, ENS, Universit\'e PSL, CNRS, Sorbonne Universit\'e, Universit\'e Paris-Diderot, Sorbonne Paris Cit\'e, Paris, France}


\begin{abstract}

A recent breakthrough in glass science has been the synthesis of  ultrastable glasses via physical vapor deposition techniques. These samples display enhanced thermodynamic, kinetic and mechanical stability, with important implications for fundamental science and technological applications. However, the vapor deposition technique is limited to atomic, polymer and organic glass-formers and is only able to produce thin film samples. Here, we propose a novel approach to generate ultrastable glassy configurations in the bulk, via random particle bonding, and using computer simulations we show that this method does indeed allow for the production of ultrastable glasses. Our technique is in principle applicable to any molecular or soft matter system, such as colloidal particles with tunable bonding interactions, thus opening the way to the design of a large class of ultrastable glasses.
\end{abstract}

\maketitle


\section{Introduction}

Glasses are
usually produced by slowly cooling a melt, with slower cooling leading to samples of higher stability~\cite{rodney2011modeling}.
Accessing highly stable glasses is important not only to answer fundamental questions, such as the existence of a phase transition to an ideal glass at the Kauzmann temperature~\cite{binder2011glassy,berthier2011theoretical,cammarota2022kauzmann}, but also for technological applications, such as creating glasses with exceptionally high strength and hardness~\cite{aji2013ultrastrong,magagnosc2016isochemical} or with strongly reduced energy dissipation~\cite{queen2013excess}.
However, in practice the range of accessible cooling rates is 
quite limited both in experiments and (even more) in numerical simulations, and hence one does not have much leeway to modify the properties of the glass since these depend in a logarithmic manner on the quench rate~\cite{binder2011glassy}. 
About a decade ago, a breakthrough was achieved by Ediger and coworkers, who were able to use vapor deposition to generate thin samples of organic,
atomic, and polymer glasses with exceptional kinetic, thermodynamic, and mechanical stability~\cite{swallen2007organic,queen2013excess,yu2013ultrastable,yoon2018testing,raegen2020ultrastable,ediger2017perspective,rodriguez2022ultrastable}.
Key to the success of this technique is the idea to use the temperature of the substrate on which the material is deposited as a tuning parameter, which has to be  optimized to maximize the ratio of the surface to bulk mobility.

Despite its attractivity, this approach does have some limitations~\cite{ediger2017perspective}.
Firstly, producing ultrastable bulk samples via vapor deposition is challenging because a slow deposition rate, of the order of $100$~microns/day, is required. 
Secondly, the produced samples are sometimes markedly anisotropic 
due to the layer-by-layer deposition process on the free surface~\cite{ediger2017perspective}.
Thirdly, the technique cannot be applied easily to colloidal glasses, because for these materials it is not easy to deposit the particles at a different temperature from that of the substrate, see e.g. Refs.~\cite{cao2017release,nguyen2020electrospray}.

Generating a well-annealed glass is a challenging task 
for computer simulations too~\cite{barrat2022computer}, a difficulty that is shared with optimization problems in computer science~\cite{kirkpatrick1983optimization}. Although many sophisticated algorithms have been proposed to tackle this problem, such as the replica-exchange Monte-Carlo (MC) method or the shoving algorithm~\cite{marinari1992simulated,hukushima1996exchange,krauth2006statistical}, these methods either
require a CPU time that depends super-linearly on the number of particles, or are too system specific. 
Recently, a particle-swap MC algorithm~\cite{grigera2001fast,gutierrez2015static}
was strongly optimized~\cite{ninarello2017models}, 
opening a new way to prepare {\it in silico} ultrastable glasses with even higher stability than experimental samples~\cite{berthier2017configurational}.
Unfortunately, however, the particle swap moves used by these algorithms are extremely hard to realize in experiments. 

\begin{figure*}[t]
\includegraphics[width=0.98\textwidth]{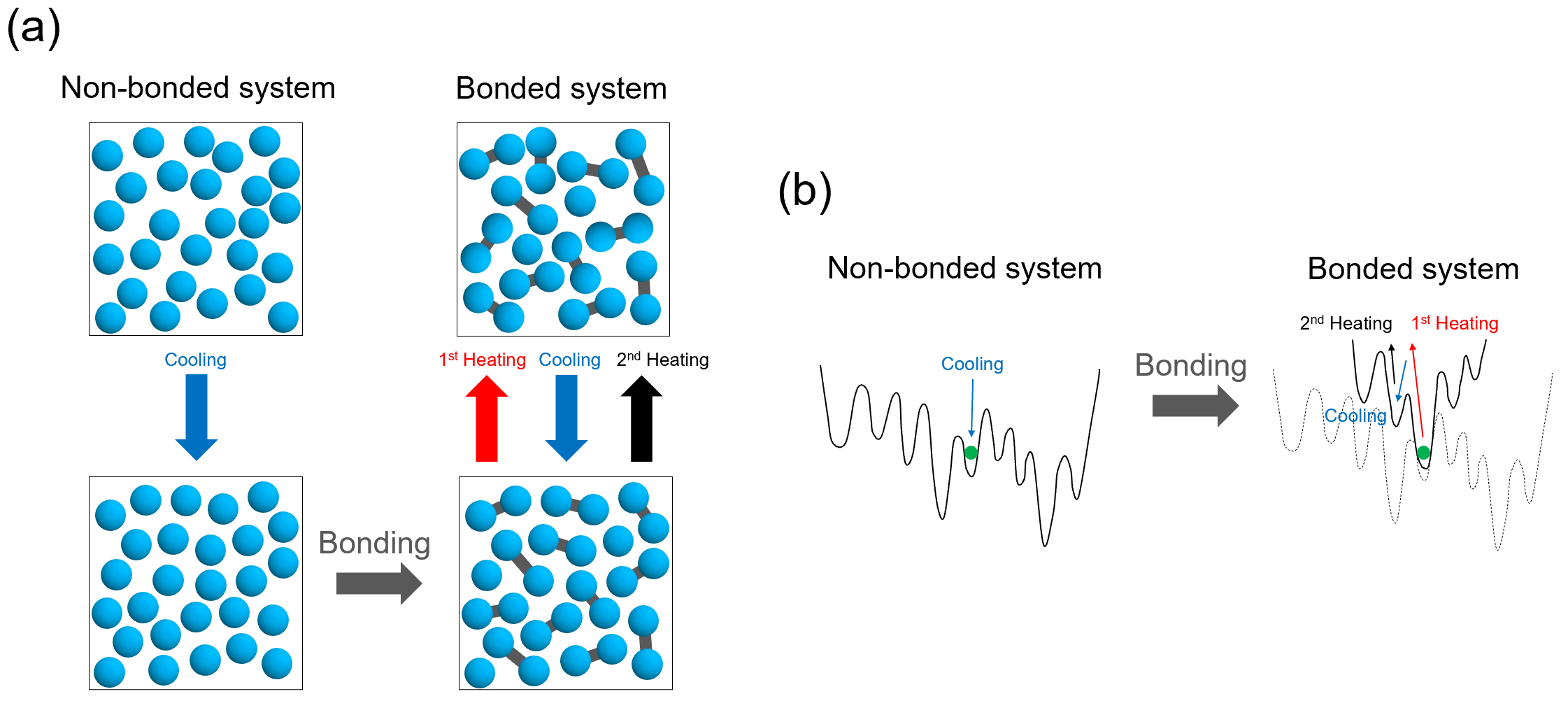}
\caption{
{\bf Schematics of the protocol.}
(a) We begin by creating an equilibrium configuration of the non-bonded (monomer) system by standard cooling to a target temperature $T$ at which the non-bonded system is still not very glassy. Note that the non-bonded system is only an auxiliary tool in our construction, while our aim is to construct an ultrastable glass of the bonded system. This is achieved by introducing the bonds via random bonding at the target temperature $T$. We now have a configuration of the bonded system, which is an ultrastable glass. To prove this, we perform heating/cooling cycles to measure kinetic stability. 
(b) Sketch of the (free) energy landscape. The non-bonded system is studied at temperatures such that barriers can be easily crossed. Once bonds are introduced, barriers become higher and the system remains trapped in a low-energy ultrastable glass basin. The first heating will allow the system to access high energy states, but subsequent cooling/heating will only permit to reach configurations that have a significantly higher energy than the one of the ultrastable glass configuration.
}
\label{fig:protocol}
\end{figure*}

An alternative approach to study the properties of ultrastable glasses is the so-called random pinning~\cite{kim2003effects,cammarota2012ideal}. After having equilibrated the liquid, one freezes permanently the positions of randomly selected particles. It can be shown that this method allows to 
access real equilibrium states of the pinned system even in the ideal glass state~\cite{kob2013probing,ozawa2015equilibrium,ozawa2018ideal},
and it can be experimentally realized in colloidal~\cite{gokhale2014growing,williams2018experimental} and molecular~\cite{kikumoto2020towards,das2021soft} systems. 
However, while random pinning
gives important insight into the thermodynamic behavior of a bulk ideal glass~\cite{cammarota2012ideal,kob2013probing,ozawa2015equilibrium,ozawa2018ideal},
it also
breaks translational invariance and as a consequence vibrational motion~\cite{angelani2018probing,shiraishi2022low}, mechanical responses~\cite{bhowmik2019effect}, and relaxation pathways~\cite{jack2013dynamical,chakrabarty2015dynamics,kob2014nonlinear} are radically altered, thus preventing one to gain insight into the dynamical properties of ultrastable glasses in the bulk.

All these approaches share the underlying idea of modifying the behavior of certain degrees of freedom
(e.g. surface mobility in vapor deposition, particle sizes in swap MC, pinned particles' movement in random pinning) by freeing or freezing them, thus allowing the glass-former to equilibrate quickly when the degrees of freedom are freed and obtain an enhanced stability of the system when they are subsequently frozen~\cite{kapteijns2019fast,hagh2022transient}.
In other words, this amounts to tune the height of the barriers in the energy landscape of the system by introducing or removing constraints on some degrees of freedom (Fig.~\ref{fig:protocol}).
Note that if one wants to produce equilibrium configurations of the frozen system, it is extremely important to perform a quiet freezing~\cite{krzakala2009hiding} of these selected degrees of freedom, i.e.,~a process that preserves the equilibrium measure~\cite{scheidler2004relaxation,cammarota2012ideal,kob2013probing,ozawa2015equilibrium}, or else the glass will be affected by out-of-equilibrium effects.

In this work,
we overcome several of the problems mentioned above, by proposing a random bonding approach, in which one freezes the distance between a subset of neighboring particle pairs (Fig.~\ref{fig:protocol}). 
Random bonding is mathematically similar to random pinning, and we will show numerically that it satisfies the requirement of quiet freezing, hence producing configurations of the bonded system that are close to equilibrium, yet without breaking the translational invariance contrarily to random pinning.
Hence, this approach allows to generate equilibrium glass configurations of a bulk system of molecules (e.g., dimers, trimers, polymers) with widely tunable stability. 
Most importantly, this method can be implemented in real experimental systems such as colloids with patches~\cite{chen2011directed,chen2012janus,wang2012colloids,duguet2011design} or DNA-linkers~\cite{mcmullen2018freely,yuan2016synthesis,mcmullen2021dna} and polymers~\cite{corezzi2002bond,corezzi2012chemical}, i.e., systems for which the interactions between the constituent particles can be modified via external parameters, see below for details.

Here, after having introduced this novel method, we present results from computer simulations that demonstrate the effectiveness of random bonding by determining for a simple glass former the kinetic stability via heating-cooling cycles and mechanical stability via stress-strain measurements.
Furthermore, we show that the generated configurations are close to equilibrium, such that no aging is detected within numerical precision. Finally, we discuss the experimental feasibility of the method.

\section{Results}

{\bf Protocol:} 
Our protocol to generate ultrastable glass configurations of the bonded system goes as follows (Fig.~\ref{fig:protocol} and Methods):
\begin{enumerate}
    \item
We first prepare, by standard cooling from high temperature, an equilibrium configuration of the
non-bonded (monomer) system at the target temperature $T$.
\item
Subsequently we pick at random two particles that are nearest neighbors, and permanently fix their distance, thus creating a dimer. 
We repeat this process, thus creating a  mixture of monomers and dimers that forms the ultrastable glass. The process stops when we reach the target dimer concentration
$c$.
\item
This mixture can now be studied via molecular dynamics (MD) simulations 
to study its thermal and mechanical response.
\end{enumerate}
The target values of temperature $T$ and dimer concentration $c$ can be tuned to obtain a glass with the desired stability. 

Before proceeding to characterize our glasses, we stress once more that the non-bonded system serves only as an auxiliary preparation tool (Fig.~\ref{fig:protocol}). The ultrastable glass we are investigating in the following is created via the random bonding process (which, in our scheme, would be akin to a vapor deposition or a random pinning) with target temperature $T=0.42$ (at which the non-bonded system has already a sluggish dynamics~\cite{coslovich2018dynamic})
and dimer concentration $c=0.95$ (thus keeping 5\% of monomers).
Subsequently we will discuss a phase diagram
of stability in the $(T,c)$ plane.

\begin{figure}[t]
\includegraphics[width=0.95\columnwidth]{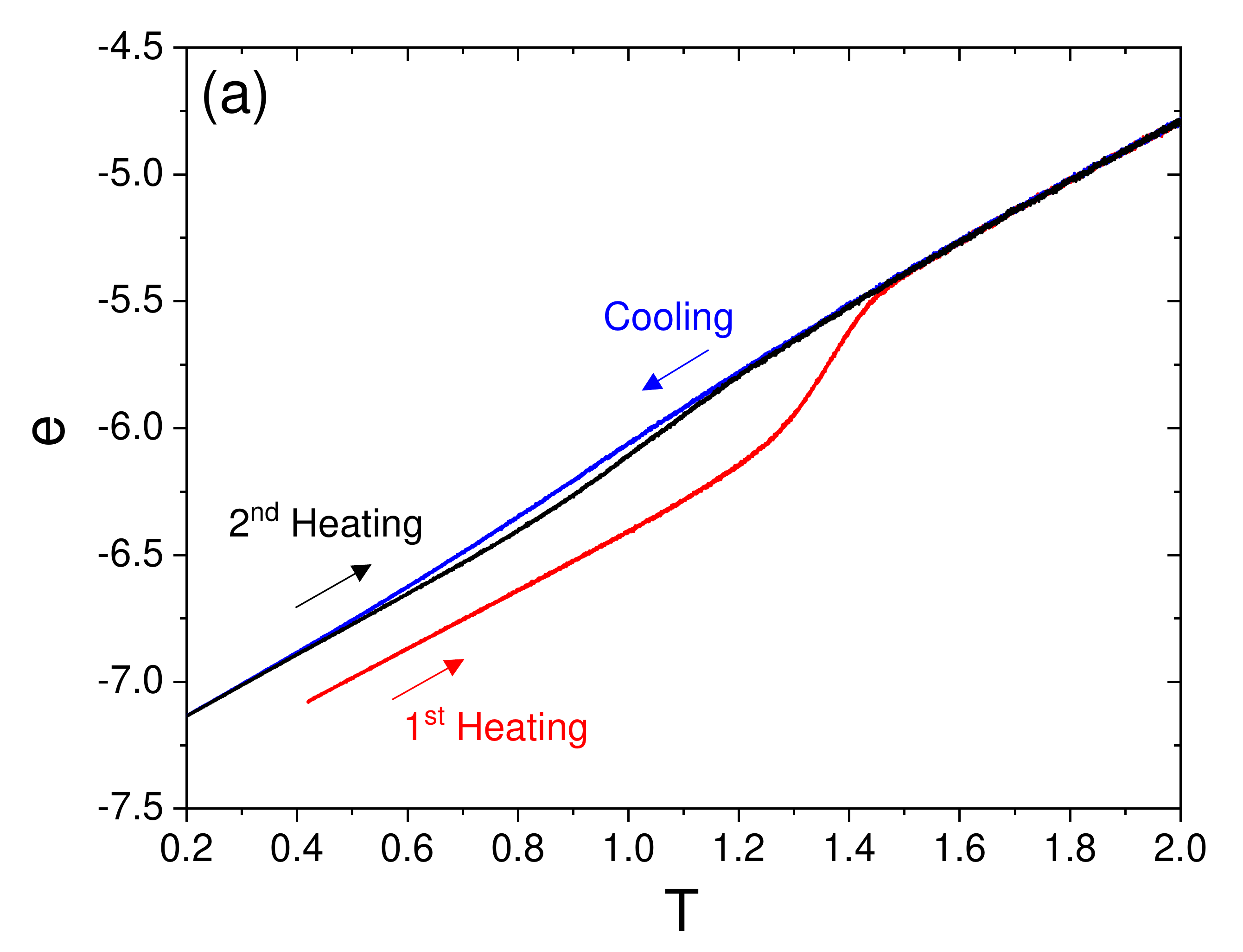}
\includegraphics[width=0.95\columnwidth]{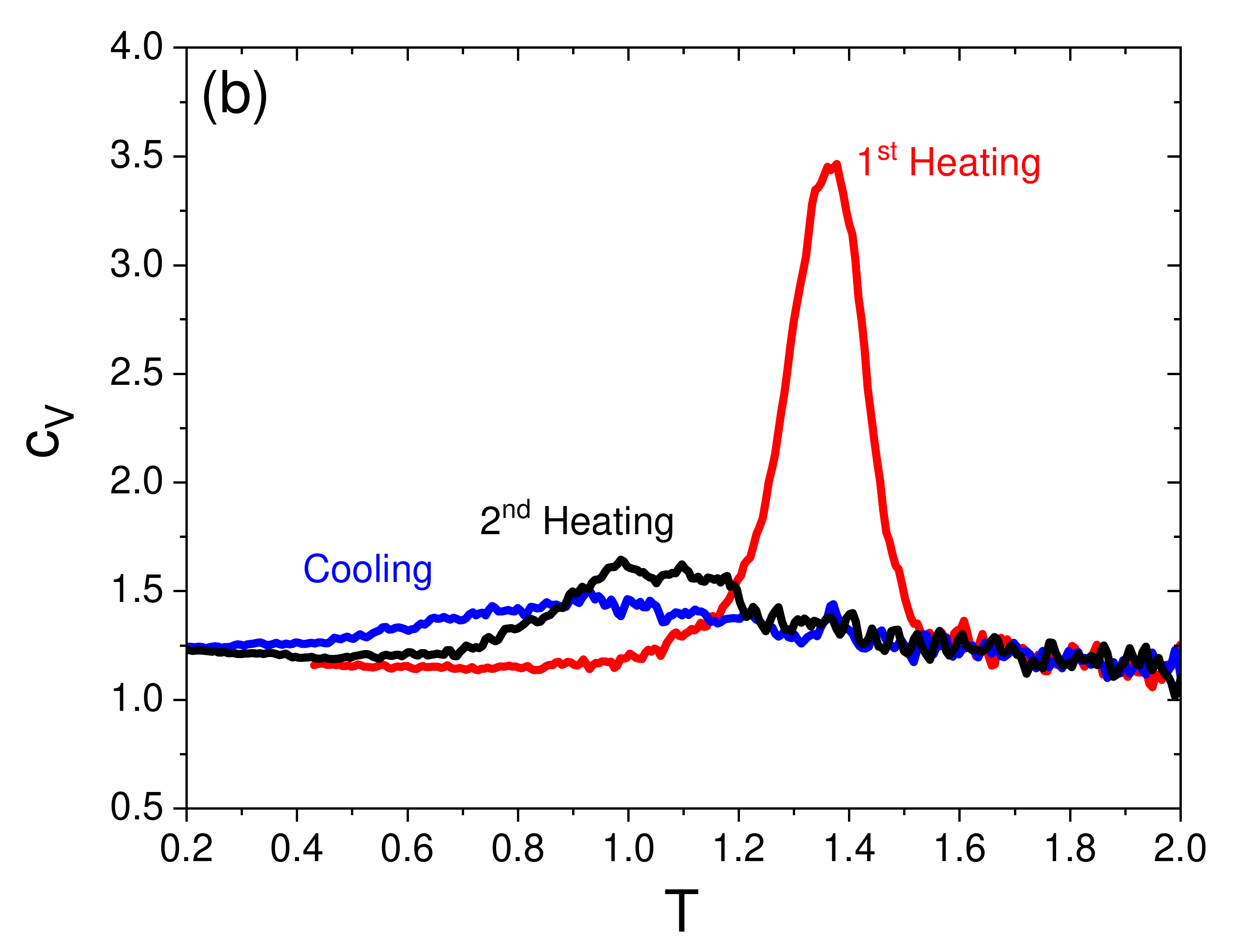}
\caption{
{\bf Heating-cooling process.} (a) The potential energy per particle  $e$ during the heating-cooling cyclic process with $K \simeq 6 \times 10^{-3}$. The curves are obtained by averaging over five different samples. The system size is $N=32400$.
(b) Specific heat $c_V$ obtained from $c_V = \mathrm{d} e/\mathrm{d} T$ using the data in (a).
}
\label{fig:stability}
\end{figure}

\vspace{0.5cm}
{\bf Kinetic stability:}
The kinetic stability of a glass-former is usually determined by a heating-cooling cyclic process \cite{swallen2007organic,hocky2014equilibrium,staley2015cooling}.
We start from the configuration created by our random bonding protocol at $T=0.42$, and then melt the glass by heating the system (keeping the bonds frozen) up to a high temperature, $T=2.1$, using a constant rate $K=|\frac{\mathrm{d}T}{\mathrm{d}t}| \simeq 6 \times 10^{-3}$.
Subsequently we cool the system down to very low $T$ ($T=0.001$), and we then heat it up a second time, using the same cooling/heating rate, $K \simeq 6\times 10^{-3}$.
Figure~\ref{fig:stability} shows the temperature evolution of the potential energy per particle and of the specific heat of the system.
We find a very pronounced hysteresis, characteristic of ultrastable glasses~\cite{swallen2007organic,hocky2014equilibrium}. During the first heating, our ultrastable glass remains in the glass state up to its kinetic glass melting (or devitrification) temperature $T_{m}^{usg} \approx 1.4$.
When the system is subsequently cooled down, it remains in the liquid state down to its normal glass transition temperature, which is estimated to be around $T_g\approx 0.8$, a value obtained from the $T$-dependence of the specific heat, Fig.~\ref{fig:stability}b. Note that this simple cooling is not able to reach an ultrastable glass state, as shown by the higher value of the energy (Fig.~\ref{fig:stability}a), and the hysteresis cycle obtained by re-heating the sample with the same heating rate (the second heating) is much smaller, indicating that the glass produced via a standard cooling procedure is indeed kinetically much less stable than the original ultrastable glass.
Figure~\ref{fig:stability}b shows the corresponding specific heat $c_V$, which makes the hysteresis even more visible.
We have checked that similar results are obtained using smaller rate $K$ (see SI). 
We thus conclude that our bonding protocol has indeed created a glass-former 
with an extremely high kinetic stability, compared to what can be achieved by simple cooling of the bonded liquid from high temperature.

\vspace{0.5cm}
{\bf Mechanical stability:}
We examine the mechanical stability of our bonded systems by following the strategy of Ref.~\cite{ozawa2018random}.
We first cool down our ultrastable glass, prepared at $T=0.42$ and $c=0.95$ as described above, to $T=0.001$ at a rate $K \simeq 6 \times 10^{-3}$, and then quench it further to zero temperature by using an energy minimization algorithm~\cite{wright1999numerical}. Subsequently, we perform a standard zero-temperature quasi-static shear simulation using Lees-Edwards boundary conditions. 
For comparison, we also consider a normally-annealed (ordinary) glass sample of the same (bonded) system, by using the configuration obtained after the first heating and cooling cycle shown in Fig.~\ref{fig:stability} (also quenched to zero temperature by energy minimization) as a starting point for the same quasi-static shear simulation.

\begin{figure}[tbp]
\includegraphics[width=0.95\columnwidth]{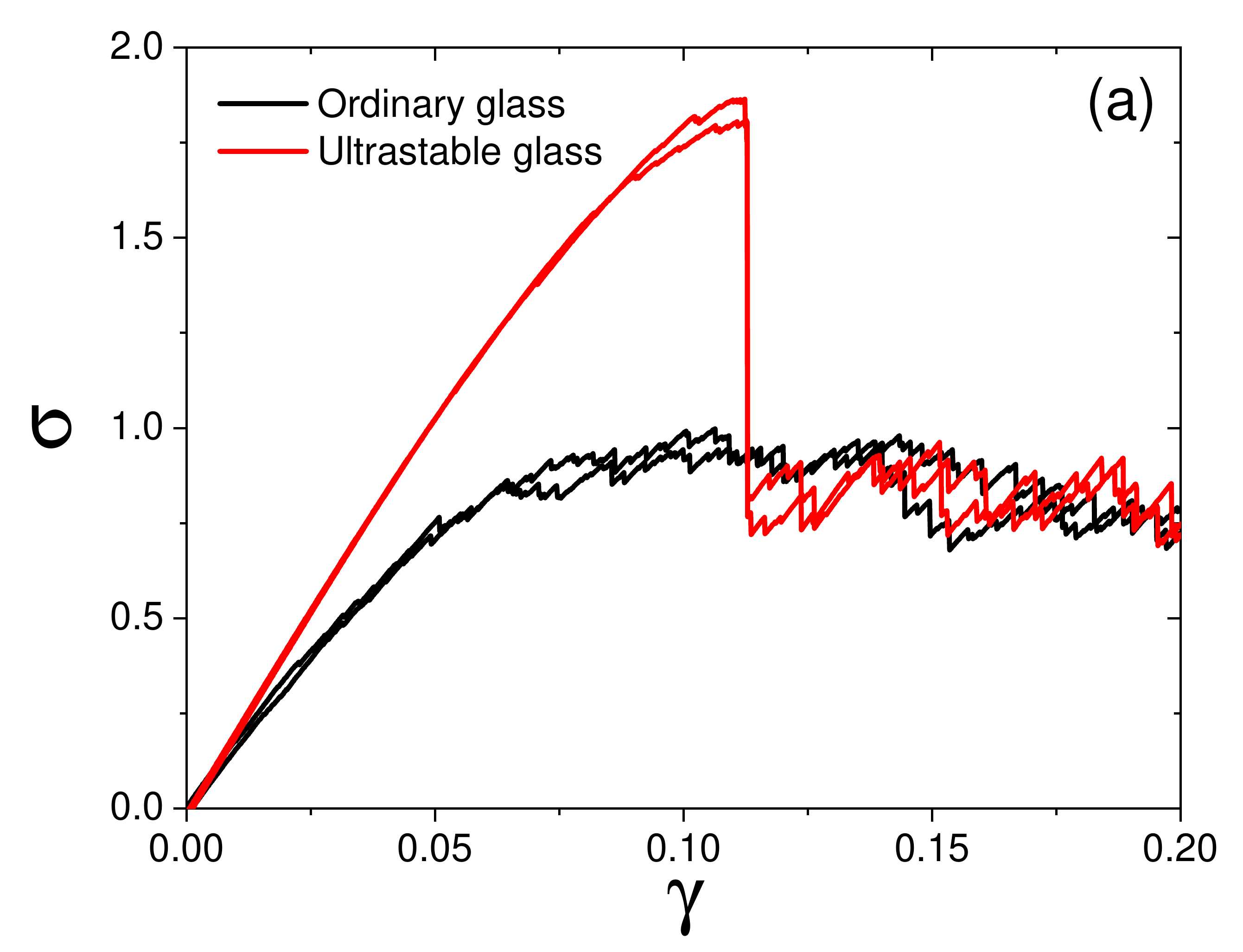}
\includegraphics[width=0.48\columnwidth]{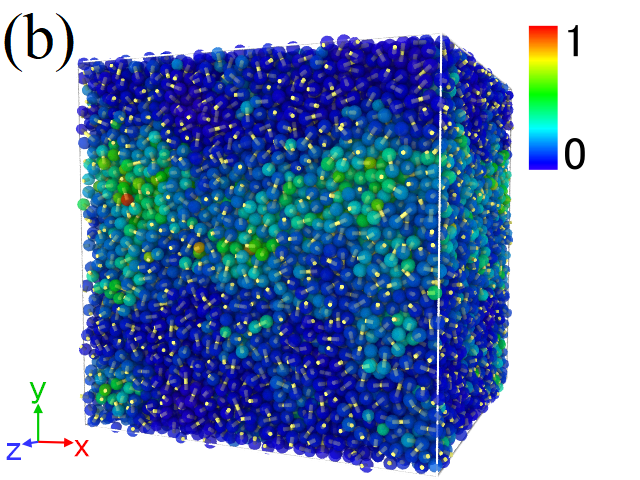}
\includegraphics[width=0.48\columnwidth]{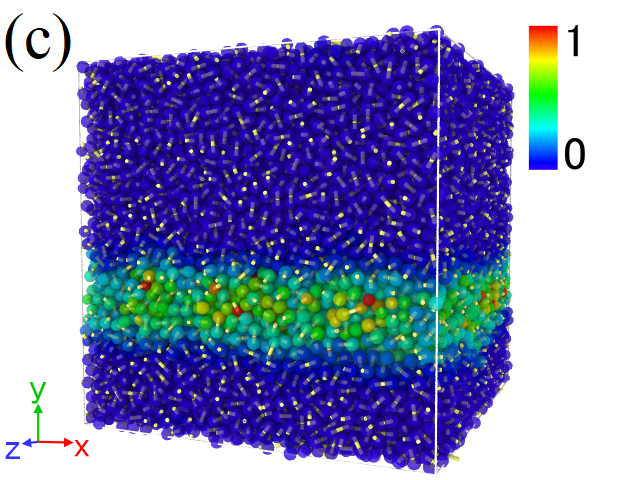}
\caption{
{\bf Mechanical test.}
(a) Stress versus strain curves of bonded glasses for ordinary and ultrastable samples.
Two independent realizations are shown, together with
snapshots of typical samples for an ordinary glass at $\gamma=0.15$ (b) and a ultrastable glass at $\gamma=0.12$ (c), respectively. The color-bar corresponds to the non-affine displacement, $D_{\rm min}^2$, measured from the origin, $\gamma=0$~\cite{falk1998dynamics}.
The bonds between particles are shown in yellow.
}
\label{fig:rheology_main}
\end{figure}

In Fig.~\ref{fig:rheology_main}a, we show the shear stress $\sigma$ as a function of strain $\gamma$ for the ultrastable and the ordinary glass samples.
The latter displays a mild stress overshoot and a gentle decrease of the stress after the maximum, indicating that the sample is ductile, at least for the system size $N=32400$ (see Refs.~\cite{richard2021finite,barlow2020ductile,rossi2022finite} for a discussion of finite size effects).
A real-space snapshot of the non-affine displacement field after the overshoot is presented in Fig.~\ref{fig:rheology_main}b, and shows a mild localization of shear. 
Remarkably, for the ultrastable sample, the stress overshoot is significantly enhanced, and an abrupt discontinuous stress drop emerges.
This brittle yielding is accompanied by a sharp system-spanning shear band, as evidenced in Fig.~\ref{fig:rheology_main}c. 
The enhanced brittleness and mechanical stability observed in Fig.~\ref{fig:rheology_main} confirms that the system is located  in a very deep local minimum of the rugged energy landscape of the bonded system, due to the preparation protocol we used, and it thus correspond to an ultrastable glass of the bonded system.

We also note that in Ref.~\cite{bhowmik2019effect} the mechanical yielding of a randomly pinned glass was studied and no brittle yielding with a system spanning shear band was observed, likely because the pinned particles break translational invariance. This demonstrates that our random bonding procedure has a huge advantage over random pinning, in that it allows to investigate the mechanical properties of realistic bulk stable glasses.

\begin{figure*}[t]
\includegraphics[width=0.66\columnwidth]{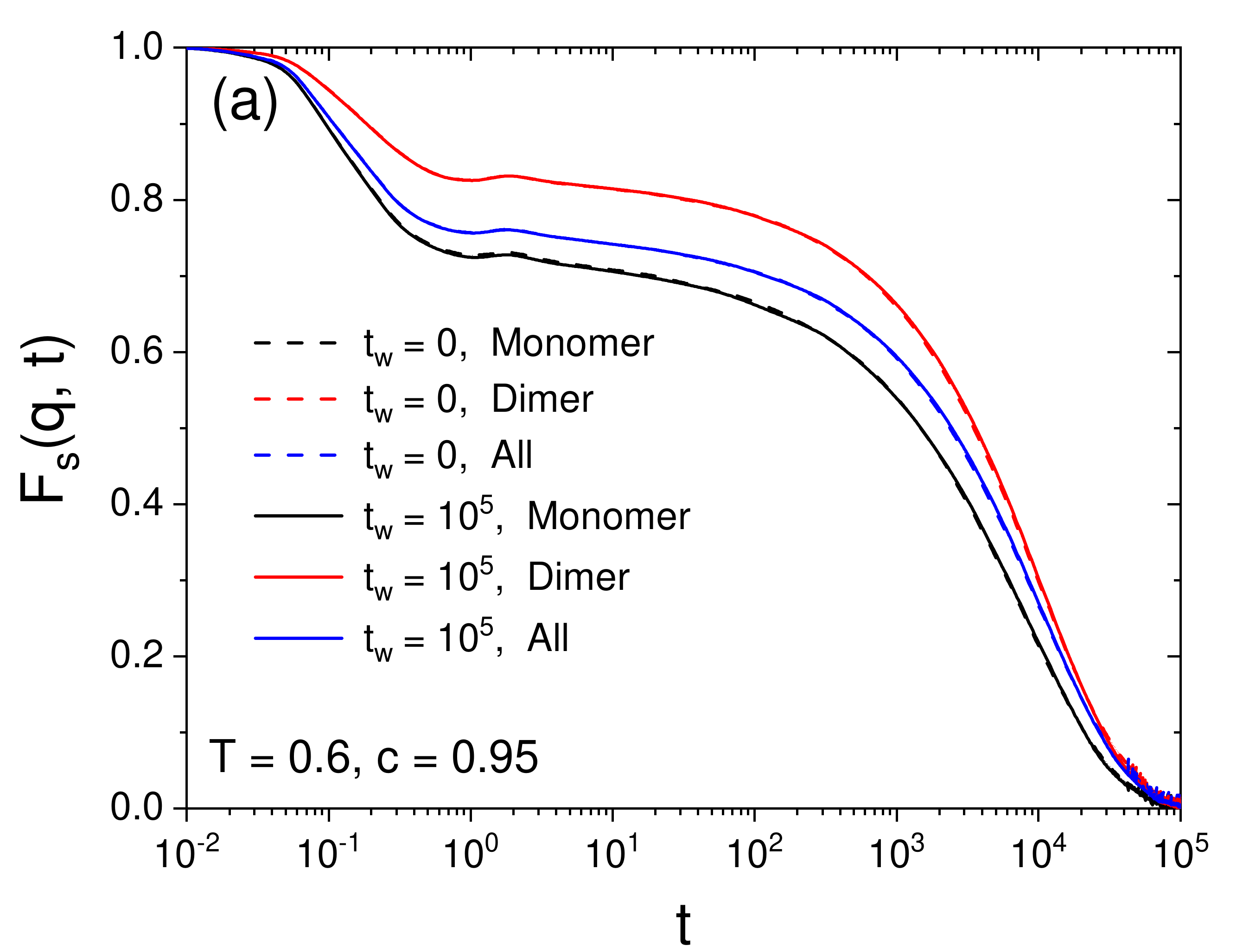}
\includegraphics[width=0.66\columnwidth]{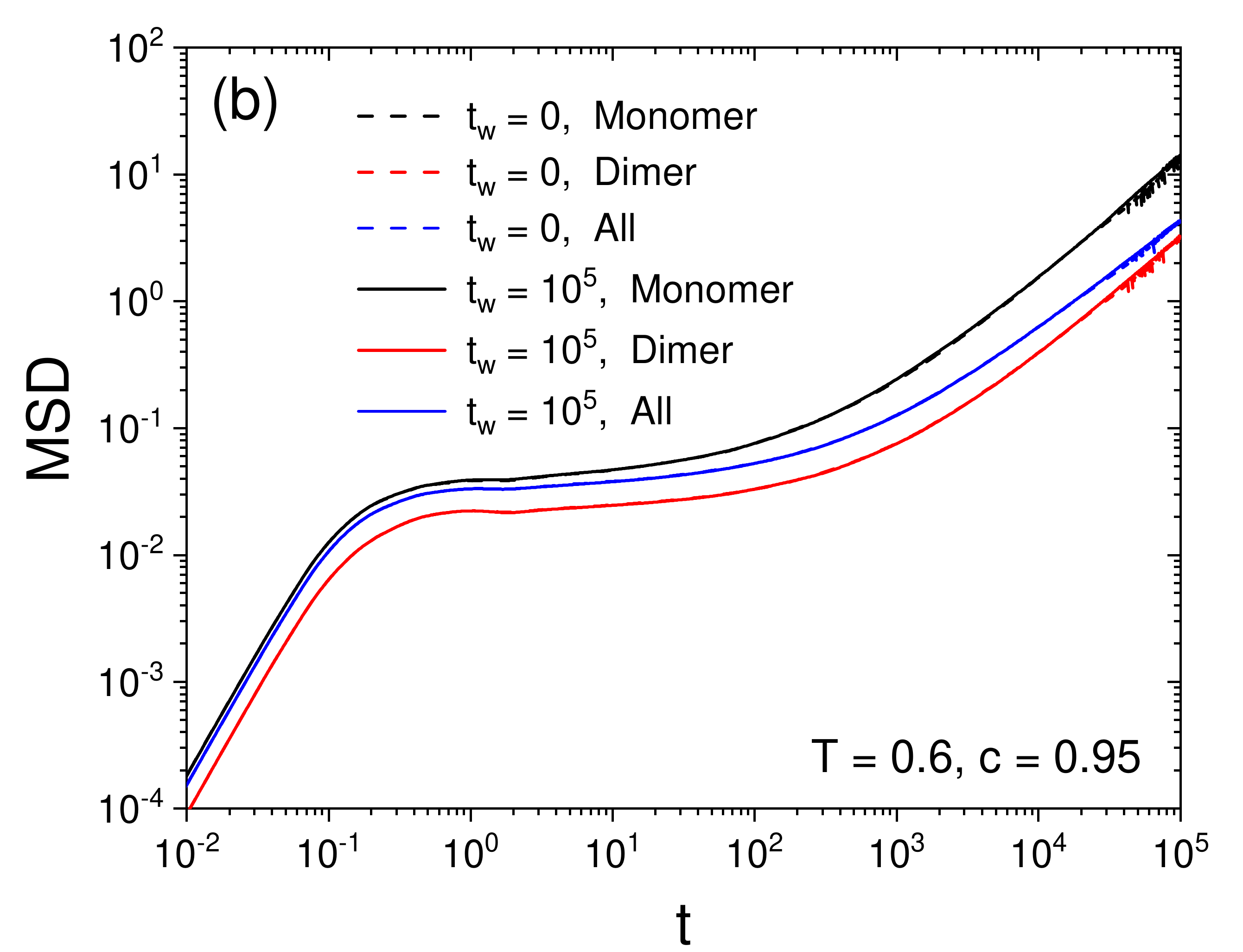}
\includegraphics[width=0.66\columnwidth]{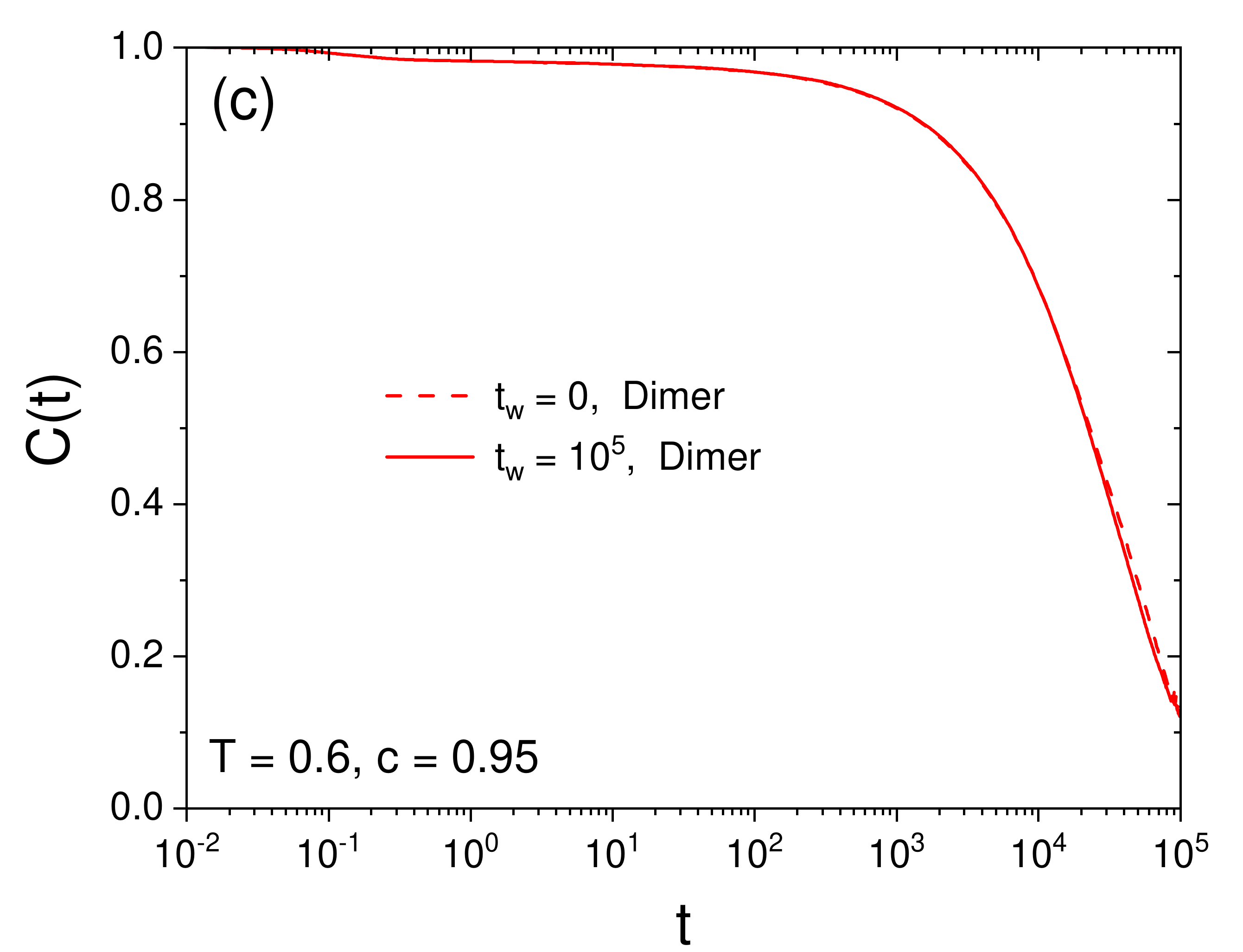}
\caption{
{\bf Absence of aging after random bonding. }
(a) Self-intermediate scattering function $F_{\rm s}(q,t)$ for $T=0.6$ and $c=0.95$ for monomers, dimers and all particles. Dashed and solid curves indicate $F_{\rm s}(q,t)$ computed from trajectories with the waiting time after the creation of the bonds $t_{\rm w}=0$ and $t_{\rm w}=10^5$, respectively.
(b) The corresponding mean-squared displacement.
(c) The corresponding rotational correlation function $C(t)$ for the dimer molecules.
}
\label{fig:absence_aging}
\end{figure*}

\vspace{0.5cm}
{\bf Absence of aging after random bonding:}
We now show that random bonding has the same quiet freezing property as random pinning, i.e.,~the system is already in thermal equilibrium right after the construction of the bonds.
This feature thus allows to study the equilibrium dynamics even in very deep glass states, simply by running a normal MD simulation starting from the configuration obtained right after bonding. Since in the present case this property holds only approximately, a more complete discussion of its validity will be given elsewhere.

To check equilibration,
we chose a target temperature of $T=0.6$ and a target dimer concentration of $c=0.95$, such that the dynamics of the bonded system is slow, but not fully arrested on the simulation time scale.
In Fig.~\ref{fig:absence_aging}a, we show
the self-part of the intermediate scattering function, $F_{\rm s}(q,t)$, defined for monomers, dimers, and all particles (see SI for the exact definitions). The wave-number $q$ is 7.25, near the location of the main peak in the static structure factor.
We observe full relaxation within the time window considered, and most importantly, that the correlation function measured right after the system is prepared ($t_{\rm w}=0$) and after a starting time $t_{\rm w}=10^5$ are numerically indistinguishable, which confirms that the bonding protocols produces equilibrated configurations, at least within numerical precision.
We confirmed the same behavior for other observables, and the mean-squared displacement and the rotational correlation function for the dimers (see SI for details) are presented in Figs.~\ref{fig:absence_aging}b and~\ref{fig:absence_aging}c, respectively, consolidating the absence of aging.

\begin{figure*}[t]
\includegraphics[width=0.66\columnwidth]{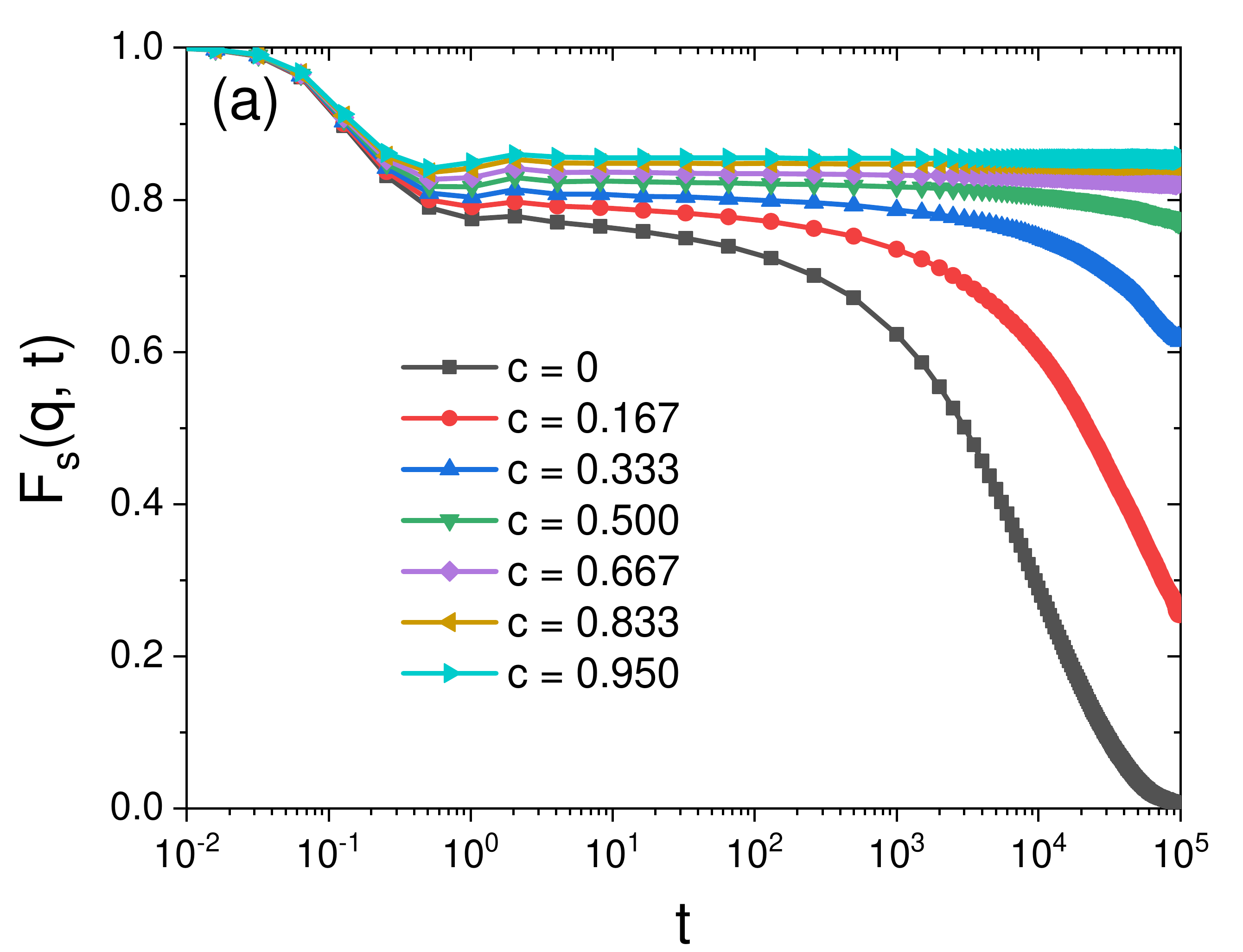}
\includegraphics[width=0.66\columnwidth]{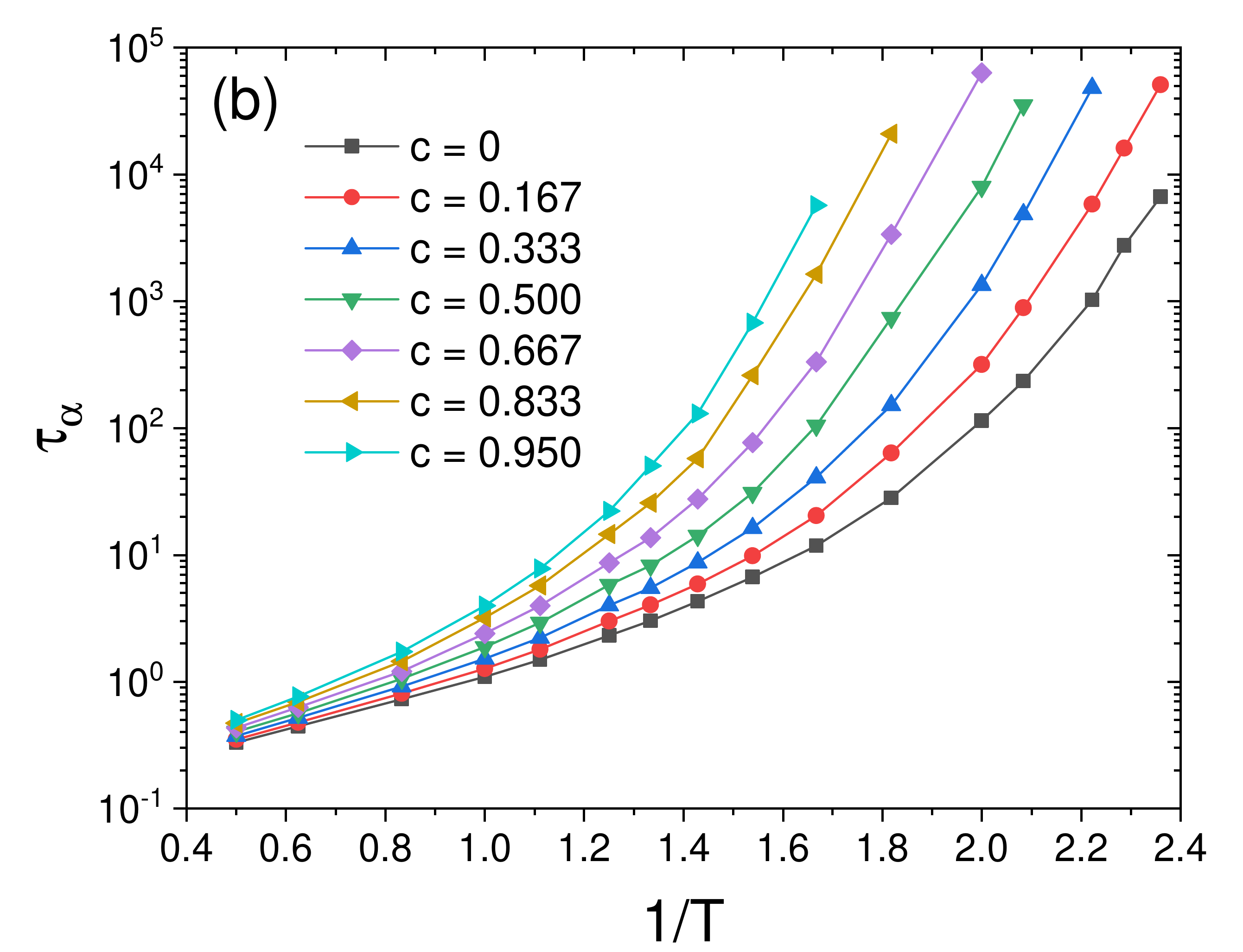}
\includegraphics[width=0.66\columnwidth]{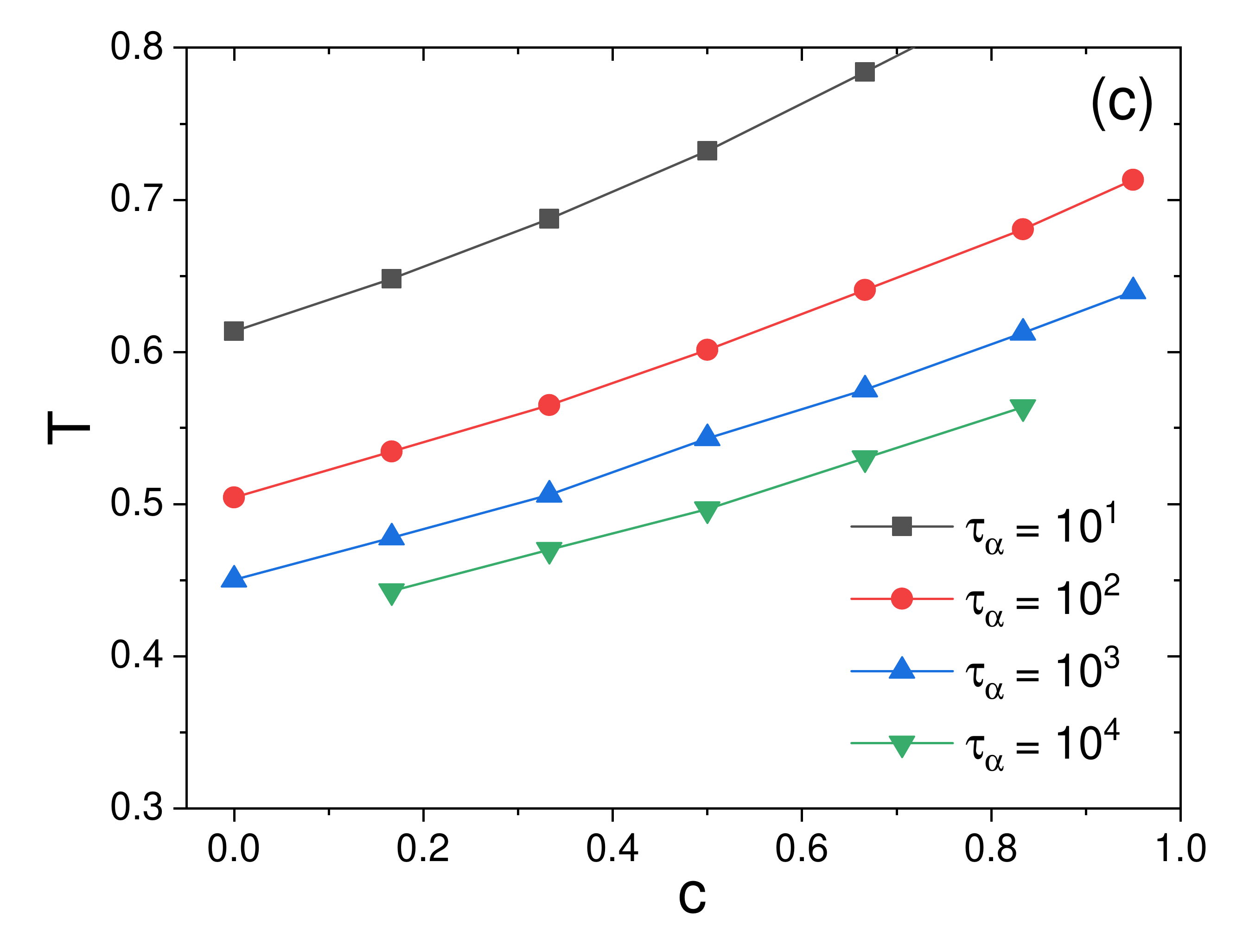}
\caption{
{\bf Slow dynamics by random bonding. }
(a) Intermediate scattering function at $T=0.4238$ and different concentrations $c$. 
The system size is $N=1200$.
(b) Structural relaxation time, $\tau_{\alpha}$, as a function of the inverse temperature, $1/T$, for several fixed concentrations $c$.
(c) Lines of constant $\tau_\alpha$ in the $(T,c)$ plane, deduced from the data in panel b.
}
\label{fig:EQ_dynamics}
\end{figure*}

\vspace{0.5cm}
{\bf Equilibrium dynamics:}
We conclude our study by investigating systematically the equilibrium dynamics in the $(T,c)$ plane.
In Fig.~\ref{fig:EQ_dynamics}a, we show $F_{\rm s}(q,t)=F^{\rm All}_{\rm s}(q,t)$ measured for all particles at $q=7.25$, after preparation of the bonded system at the target temperature $T=0.4238$ for different concentrations $c$.  (See the SI for corresponding results for other values of $T$.)
As argued above, this corresponds to the equilibrium dynamics of the bonded system.
For the original KA model, $c=0$, we observe full relaxation within the time window considered, although at this $T$ the dynamics is already glassy. The dynamics slows down very quickly as $c$ is increased and for $c>0.5$ it is completely frozen on the timescale of the simulation, demonstrating that the system has entered a truly glassy regime.
We define the structural relaxation time $\tau_{\alpha}$ via $F_{\rm s}(q,\tau_{\alpha})=1/e$ and in Fig.~\ref{fig:EQ_dynamics}b we show its temperature and $c$-dependence in an Arrhenius plot.
From this graph one concludes that increasing $c$ leads to a quick slowing down of the relaxation dynamics, in agreement with the result shown in Fig.~\ref{fig:EQ_dynamics}a, and that this effect is strongly enhanced if $T$ is decreased, i.e.,~a behavior that is qualitatively similar to the one found for randomly pinned systems~\cite{kob2013probing,ozawa2015equilibrium}. From these results, we can draw iso-$\tau_\alpha$ lines in the $(T,c)$ plane
that characterize the stability of the glasses prepared by random bonding at these values of target temperature and dimer concentration (Fig.~\ref{fig:EQ_dynamics}c).

Based on an extrapolation of these results, we estimate that the glass analyzed in Figs.~\ref{fig:stability} and \ref{fig:rheology_main}, prepared at ${T=0.42}$ and $c=0.95$, corresponds to an equilibrium relaxation time $\tau_{\alpha}\approx 10^{12}$, thus about a factor of $10^7$ larger
than the largest $\tau_{\alpha}$ we accessed in our simulations (see SI). This demonstrates that the bonding process does indeed allow to generate glasses with a fictive temperature that is far lower than one can access by means of standard algorithms, and comparable to that achieved by the swap algorithm~\cite{ninarello2017models}.

Finally, 
it has been reported that, in contrast to conventional glass-forming liquids, the equilibrium dynamics of pinned systems shows a decoupling between self and collective relaxation and a suppression of dynamical heterogeneity upon approaching the glass transition, due to the highly confined environment~\cite{jack2013dynamical,kob2014nonlinear}.
We confirm that the randomly-bonded systems do not show such a decoupling between the self and collective parts (see SI), as expected for a system that has translational invariance.
We leave the study of the dynamical heterogeneity of randomly-bonded systems for future work.

\section{Discussion}

We have introduced a novel protocol that allows one to prepare highly stable equilibrium glasses in the bulk (Fig.~\ref{fig:protocol}). 
The approach requires the generation of an equilibrated configuration of a simple glass-forming system at intermediate temperatures, i.e., where its dynamics is not yet arrested (which is routinely done in experiments and in simulations), and subsequently the introduction of random bonds between neighboring molecules (here monomers) to form a certain fraction of dimers.
We numerically demonstrated that the resulting configurations of the bonded system are in equilibrium, and our simulations show that their dynamics is extremely slow, indicating that one does indeed access deep glassy states.
Moreover, we have demonstrated the strong enhancement of kinetic and mechanical stability via a non-equilibrium heating-cooling process and athermal quasi-static shear simulations, respectively. 

\begin{figure}[tbp]
\includegraphics[width=0.85\columnwidth]{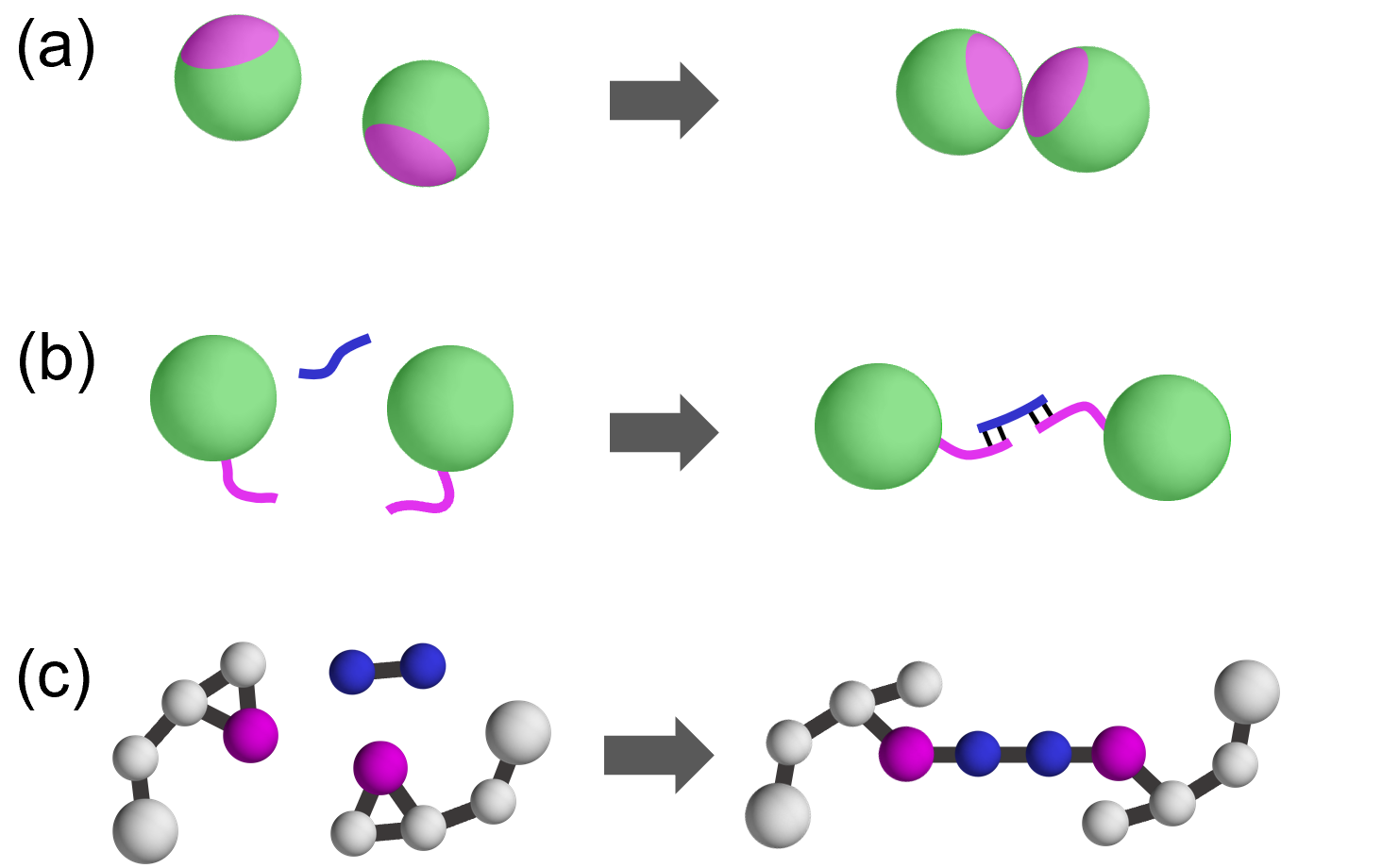}
\caption{{\bf Examples of bond formation processes in experimental systems.}
(a)~Inducing attractive interactions between patchy colloids by changing the salt concentration. (b)~Varying temperature or irradiation with ultra-violet light in colloids with DNA-linkers. (c)~Chemical bonding between epoxy groups with a curing agent. 
}
\label{fig:experiment}
\end{figure}

We emphasize that it is possible to realize the proposed random bonding in real experiments of soft matter systems, such as colloids and emulsions, for which existing vapor deposition techniques cannot be applied easily, thus allowing to produce ultrastable glasses for a new class of glass-formers. One can, e.g., tune interactions between colloidal particles, by introducing patches or  DNA-linkers~\cite{duguet2011design,mcmullen2018freely},
as schematically shown in Figs.~\ref{fig:experiment}a and \ref{fig:experiment}b. Importantly, one can turn on the bond interactions at any time by varying the concentration of salt or the temperature~\cite{mcmullen2018freely,mcmullen2021dna}, or using ultra-violet light~\cite{yuan2016synthesis}.
To demonstrate this, we also carried out experiments using two-dimensional patchy colloids. Starting from an equilibrated binary mixture of patchy and non-patchy colloids (monomers)~\cite{sciortino2011reversible}, we increase the salt concentration to activate the patchy interaction between patchy colloid pairs, leading to the formation of dimers with strong bonds. The resulting configurations have a finite concentration of dimers (with a few trimers and tetramers), demonstrating that indeed one can turn on the bond interactions at any time (see SI for details).
A further possibility to
implement random bonding in molecular or polymeric systems is schematically shown in Fig.~\ref{fig:experiment}c. For epoxy resins, e.g., one can control polymerization by adding to the sample a curing agent~\cite{corezzi2002bond,corezzi2012chemical} or exposing it to ultra-violet light~\cite{rehbein2020experimental}.
These few examples illustrate that it is in principle possible to produce ultrastable glasses by particle bonding in experiments, thus opening new research directions in the field.

The random bonding between two neighboring particles to produce dimer systems is of course only one possible implementation of our approach and it is straightforward to extend the method to construct trimers, polymers, or more complex molecules. We expect that more stable configurations can be obtained when a particle can have more bonds, due to the freezing of additional degrees of freedom~\cite{zou2009packing,hagh2022transient,nandi2021connecting}.
When the number of bonds increases, a network structure percolates through the system, leading to a glass-gel crossover~\cite{michel2000percolation,chaudhuri2015relaxation,sciortino2011reversible}.  
We expect that the random bonding will thus provide a recipe to produce ultrastable gels as well, thus allowing to study also the glass-gel crossover in equilibrium.

By exploring these ideas, it should be possible to realize an ideal glass transition in a bulk equilibrium bonded system akin to the randomly pinned systems~\cite{cammarota2012ideal,kob2013probing,ozawa2015equilibrium}, yet without breaking translational invariance. Hence we expect that our study will pave the way to an experimental realization of ideal bulk glasses.

\medskip

\section{Methods}

As a monomer system, we use the standard Kob-Andersen (KA) model in three dimensions~\cite{kob1995testing}.
The original, non-bonded (monomer) KA model is simulated in the $NVT$ ensemble with $N$ particles in a periodic volume $V$ at the target temperature $T$ (see SI) to produce the initial equilibrium configuration.
We then introduce a certain fraction of diatomic dumbbell molecules, i.e.,~dimers, by random bonding particles that have a distance less than $R_{\rm b}=1.5$.
(All quantities are expressed in standard reduced Lennard-Jones units~\cite{kob1995testing}.)
In this way, we create a
system with $N_{\rm m}$ monomers and $N_{\rm d}$ dimers. By construction, $N=N_{\rm m}+2N_{\rm d}$. The dimer concentration is
$c=\frac{2N_{\rm d}}{N}=\frac{N-N_{\rm m}}{N}$.
To simulate the bonded system at finite temperature, 
we use
 the RATTLE algorithm to keep the bond lengths fixed~\cite{andersen1983rattle}. 
To study mechanical responses, we use an athermal quasi-static shear simulation via an energy minimization algorithm~\cite{maloney2006amorphous}, in which the rigid bonds were replaced by a harmonic spring with sufficiently hard stiffness, see SI for details.

\vspace{0.3cm}
{\bf Acknowledgments:}
We thank J.-L.~Barrat, M.~Ediger, V.~F.~Hagh, H.~Ikeda, F.~C.~Mocanu and M.~A.~Ramos for discussions
and K.~Yoshihara for his contribution to the preliminary patchy particle experiment.
This project has received funding from the European Research Council (ERC) under the European Union's Horizon 2020 research and innovation programme (grant agreement n. 723955 - GlassUniversality). W.K.~is member of the Insititut universitaire de France.

\bibliography{bonding}

\beginsupplement

\centerline{\Large\bf SUPPLEMENTARY INFORMATION}

\section{STATISTICAL MECHANICS OF BONDED SYSTEMS}

We discuss here the statistical mechanics of randomly-bonded glass formers composed of monomers and dimers.
For this we randomly choose a fraction of pairs of neighboring particles from an equilibrium configuration and create bonds with a rigid-body constraint.
This section also defines some of the notation used in the main text of the manuscript.

\subsection{Variable transformation}

Consider a system of $N$ point particles in three dimensions, $d=3$, such as the Kob-Andersen model~\cite{kob1995testing}.
The Hamiltonian $H$ and the partition function $Z$ are given by
\begin{equation}
H({\bf r}^N, {\bf p}^N) = \sum_{i=1}^N \frac{{\bf p}_i^2}{2m_i} + U({\bf r}^N) \ , \qquad \qquad
Z = \int \mathrm{d}{\bf r}^N \mathrm{d}{\bf p}^N \, \exp[-\beta H({\bf r}^N, {\bf p}^N)] \ ,
\label{eq:original_H}
\end{equation}
where ${\bf r}_i$, ${\bf p}_i$, and $m_i$ are the position, momentum, and mass of the $i$-th particle, respectively. $U$ is the potential energy, and $\beta=1/T$ is the inverse temperature.
In this paper, we use a shorthand notation for a vector of $N$ variables, e.g., ${\bf r}^N=( {\bf r}_1, {\bf r}_2, ..., {\bf r}_N)$.
Note that we omit the combinatorial factors such as $N!$ and the Planck's constant $h$, because we will not discuss the absolute value of the free energy or entropy in this work.

\begin{figure}[htbp]
\includegraphics[width=0.5\columnwidth]{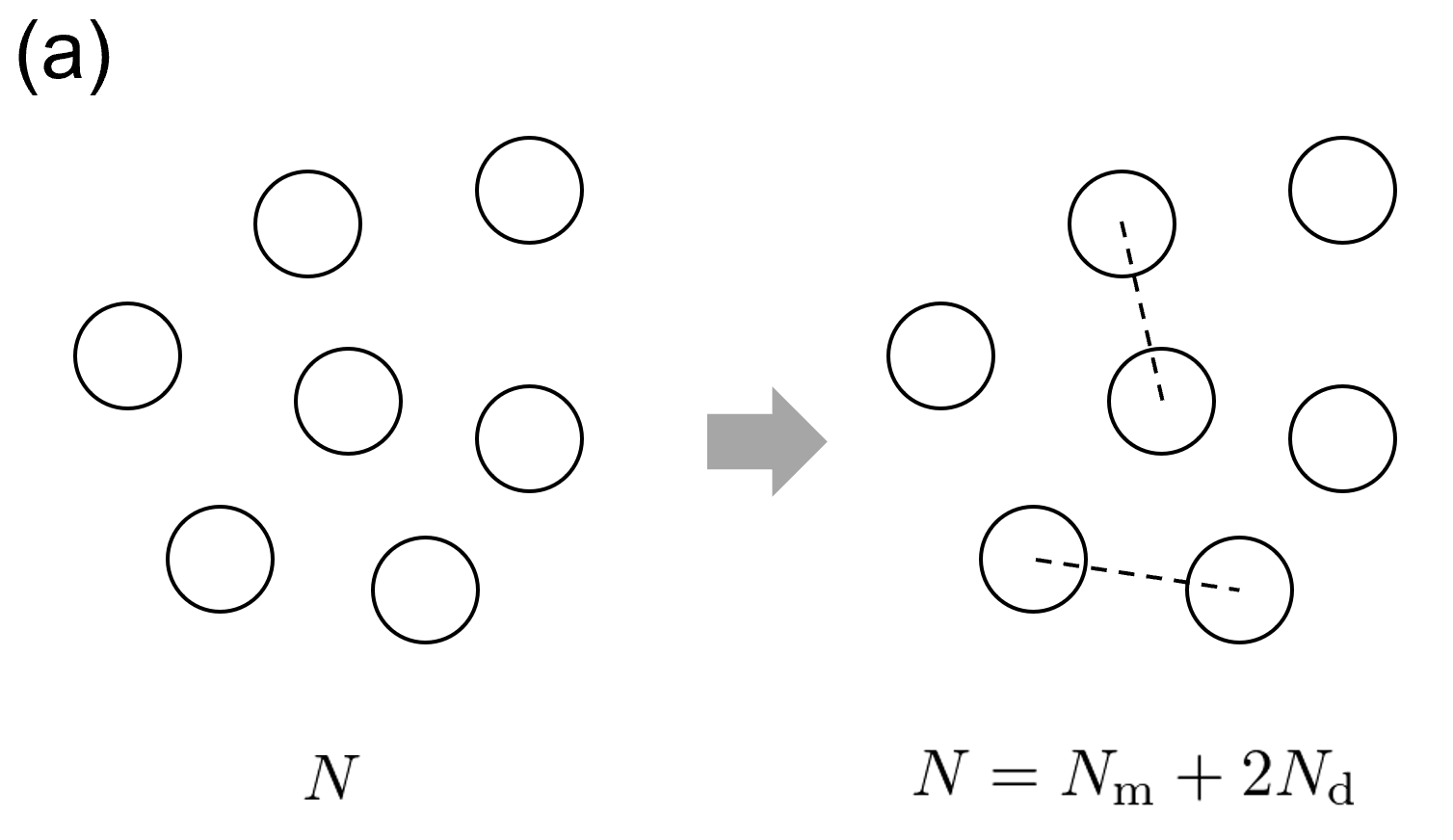}
\qquad \qquad
\includegraphics[width=0.3\columnwidth]{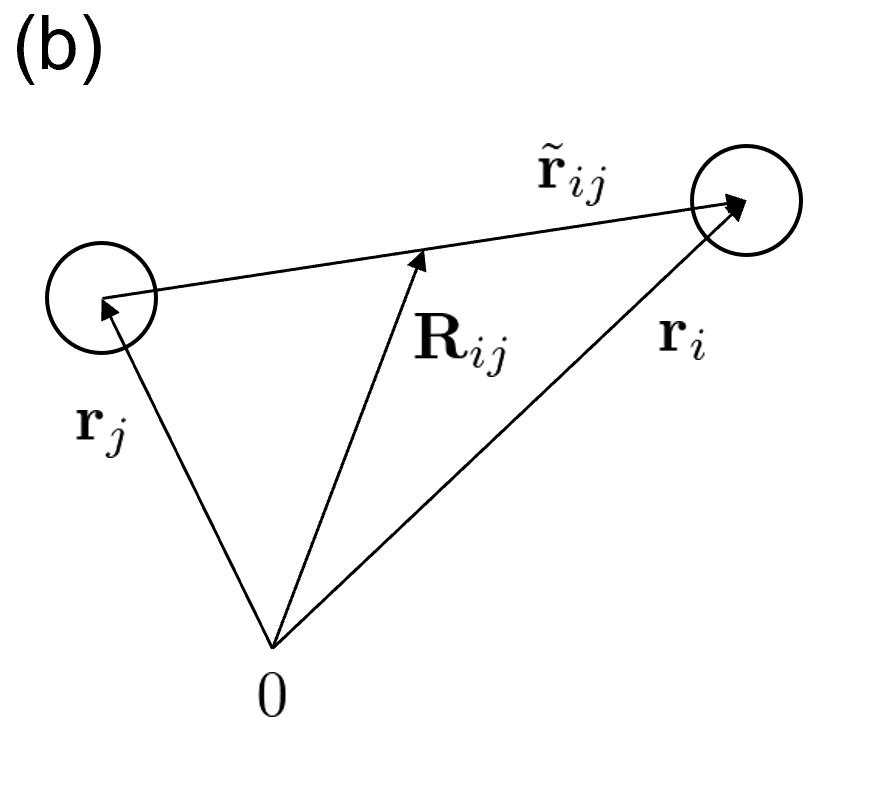}
\caption{(a): Schematic plot for making virtual bonds. (b): The center of mass and relative position describe a dimer connecting particles $i$ and $j$.}
\label{fig:virtual}
\end{figure}

We consider the following variable transformation from Cartesian coordinates to Jacobi coordinates. We assign virtual bonds between two neighboring particles chosen randomly from an equilibrium configuration, as schematically shown in Fig.~\ref{fig:virtual}a.
Note that this is a virtual operation, and the actual system is not altered at all, i.e., this is merely a variable transformation.
After the transformation, $N$ particles are classified into the group of $N_{\rm m}$ monomers and the others into the group of $N_{\rm d}$ dimers.
By construction, $N=N_{\rm m}+2N_{\rm d}$.
The index of the particle associated with the monomers and dimers belongs to sets $i \in \mathcal{M}$ and $i \in \mathcal{D}$, respectively.

For the monomers, we continue to use the Cartesian coordinates.
For the dimers, instead, we use the Jacobi coordinates for the two-body problem, using the center of mass of two particles and relative position as shown in Fig.~\ref{fig:virtual}b.
The center of mass and relative position for a dimer composed of the $i$-th and $j$-th particles are given by ${\bf R}_{ij}=\frac{m_i {\bf r}_i + m_j {\bf r}_j}{m_i+m_j}$ and $\tilde{{\bf r}}_{ij}={\bf r}_i-{\bf r}_j$, respectively.
The corresponding momenta are denoted by ${\bf P}_{ij}$ and $\tilde{{\bf p}}_{ij}$, respectively.
Also, the total mass and the reduced mass are given by $M_{ij}=m_i+m_j$ and $\mu_{ij}=\frac{m_i m_j}{m_i+m_j}$, respectively.
The described variable transformation is formally written by $({\bf r}^N, {\bf p}^N) \to ( {\bf r}^{N_{\rm m}}, {\bf p}^{N_{\rm m}}, {\bf R}^{N_{\rm d}}, {\bf P}^{N_{\rm d}}, \tilde{{\bf r}}^{N_{\rm d}}, \tilde{{\bf p}}^{N_{\rm d}} )$.

Under this variable transformation, we can rewrite the Hamiltonian $H$ as,
\begin{equation}
H( {\bf r}^{N_{\rm m}}, {\bf p}^{N_{\rm m}}, {\bf R}^{N_{\rm d}}, {\bf P}^{N_{\rm d}}, \tilde{{\bf r}}^{N_{\rm d}}, \tilde{{\bf p}}^{N_{\rm d}} ) = \sum_{i \in \mathcal{M}} \frac{{\bf p}_i^2}{2 m_i} + \sum_{\substack{i,j \in \mathcal{D} \\ (i<j)}} \left( \frac{{\bf P}_{ij}^2}{2 M_{ij}} + \frac{\tilde{{\bf p}}_{ij}^2}{2 \mu_{ij}} \right) + U( {\bf r}^{N_{\rm m}}, {\bf R}^{N_{\rm d}}, \tilde{{\bf r}}^{N_{\rm d}} ) \ ,
\label{eq:H_Jacobi}
\end{equation}
and the partition function $Z$ as
\begin{equation}
Z = \int(\prod_{i \in \mathcal{M}} \mathrm{d}{\bf r}_i \mathrm{d}{\bf p}_i) \int (\prod_{\substack{i,j \in \mathcal{D} \\ (i<j)}} \mathrm{d} {\bf R}_{ij}\mathrm{d} {\bf P}_{ij}) \int (\prod_{\substack{i,j \in \mathcal{D} \\ (i<j)}} \mathrm{d} \tilde{{\bf r}}_{ij}\mathrm{d} \tilde{{\bf p}}_{ij}) \exp[-\beta H( {\bf r}^{N_{\rm m}}, {\bf p}^{N_{\rm m}}, {\bf R}^{N_{\rm d}}, {\bf P}^{N_{\rm d}}, \tilde{{\bf r}}^{N_{\rm d}}, \tilde{{\bf p}}^{N_{\rm d}} )] \ .
\end{equation}

We further proceed with the variable transformation for the relative movements of dimers by using spherical coordinates in $d=3$, which is formally written as
$(\tilde{\bf r}^{N_{\rm d}}, \tilde{\bf p}^{N_{\rm d}}) \to (\tilde{r}^{N_{\rm d}}, \theta^{N_{\rm d}}, \varphi^{N_{\rm d}}, p_{\tilde{r}}^{N_{\rm d}}, p_{\theta}^{N_{\rm d}}, p_{\varphi}^{N_{\rm d}})$.
The Liouville's theorem ensures that the Jacobian is unity for this transformation, namely, $\mathrm{d} \tilde{{\bf r}}_{ij}\mathrm{d} \tilde{{\bf p}}_{ij}=\mathrm{d} \tilde{r}_{ij} \mathrm{d}\theta_{ij}\mathrm{d}\varphi_{ij} \mathrm{d}p_{\tilde{r}_{ij}} \mathrm{d}p_{\theta_{ij}} \mathrm{d}p_{\varphi_{ij}}$.
Also, the kinetic part of the Hamiltonian can be written as
\begin{equation}
\frac{\tilde{\bf p}_{ij}^2}{2 \mu_{ij}} = \frac{p_{\tilde{r}_{ij}}^2}{2 \mu_{ij}} + \frac{p_{\theta_{ij}}^2}{2I_{ij}} + \frac{p_{\varphi_{ij}}^2}{2 I_{ij} \sin^2\theta_{ij}} \ ,
\label{eq:kinetic_spherical}
\end{equation}
where $I_{ij}=\mu_{ij}\tilde{r}_{ij}^2$ is the moment of inertia, $p_{\tilde{r}_{ij}} = \mu_{ij} \dot{ \tilde{r}}_{ij}$, $p_{\theta_{ij}} = I_{ij} \dot{\theta}_{ij}$, and $p_{\varphi_{ij}} = I_{ij} (\sin^2\theta_{ij}) \dot{\varphi}_{ij}
$ are new momenta.
We note that as one can see in Eq.~(\ref{eq:kinetic_spherical}), the kinetic term also depends on the coordinates.
Thus we cannot treat the momenta and positions separately, unlike in the random pinning setting.

We finally arrive at the expressions for the Hamiltonian $H$, the partition function $Z$ and the Boltzmann distribution $\rho = \exp[-\beta H]/Z$
that are suitable for our study:
\begin{eqnarray}
H( {\bf r}^{N_{\rm m}}, {\bf p}^{N_{\rm m}}, {\bf R}^{N_{\rm d}}, {\bf P}^{N_{\rm d}}, \tilde{r}^{N_{\rm d}}, \theta^{N_{\rm d}}, \varphi^{N_{\rm d}}, p_{\tilde{r}}^{N_{\rm d}}, p_{\theta}^{N_{\rm d}}, p_{\varphi}^{N_{\rm d}} ) = \qquad \qquad \qquad \qquad \qquad \qquad \qquad \nonumber \\ \sum_{i \in \mathcal{M}} \frac{{\bf p}_i^2}{2 m_i} + \sum_{\substack{i,j \in \mathcal{D} \\ (i<j)}} \left( \frac{{\bf P}_{ij}^2}{2 M_{ij}} + \frac{p_{\tilde{r}_{ij}}^2}{2 \mu_{ij}} + \frac{p_{\theta_{ij}}^2}{2I_{ij}} + \frac{p_{\varphi_{ij}}^2}{2 I_{ij}\sin^2\theta_{ij}} \right) + U( {\bf r}^{N_{\rm m}}, {\bf R}^{N_{\rm d}}, \tilde{r}^{N_{\rm d}}, \theta^{N_{\rm d}}, \varphi^{N_{\rm d}} ) \ ,
\end{eqnarray}
\begin{eqnarray}
Z = \int(\prod_{i \in \mathcal{M}} \mathrm{d}{\bf r}_i \mathrm{d}{\bf p}_i) \int (\prod_{\substack{i,j \in \mathcal{D} \\ (i<j)}} \mathrm{d} {\bf R}_{ij}\mathrm{d} {\bf P}_{ij}) \int (\prod_{\substack{i,j \in \mathcal{D} \\ (i<j)}} \mathrm{d} \tilde{r}_{ij} \mathrm{d}\theta_{ij}\mathrm{d}\varphi_{ij} \mathrm{d}p_{\tilde{r}_{ij}} \mathrm{d}p_{\theta_{ij}} \mathrm{d}p_{\varphi_{ij}}) \nonumber \\ \times \exp[-\beta H( {\bf r}^{N_{\rm m}}, {\bf p}^{N_{\rm m}}, {\bf R}^{N_{\rm d}}, {\bf P}^{N_{\rm d}}, \tilde{r}^{N_{\rm d}}, \theta^{N_{\rm d}}, \varphi^{N_{\rm d}}, p_{\tilde{r}}^{N_{\rm d}}, p_{\theta}^{N_{\rm d}}, p_{\varphi}^{N_{\rm d}} )] \ .
\end{eqnarray}
We define the average over the total Boltzmann distribution, $\langle \cdots \rangle$, by
\begin{eqnarray}
\langle \cdots \rangle &=& \int(\prod_{i \in \mathcal{M}} \mathrm{d}{\bf r}_i \mathrm{d}{\bf p}_i) \int (\prod_{\substack{i,j \in \mathcal{D} \\ (i<j)}} \mathrm{d} {\bf R}_{ij}\mathrm{d} {\bf P}_{ij}) \int (\prod_{\substack{i,j \in \mathcal{D} \\ (i<j)}} \mathrm{d} \tilde{r}_{ij} \mathrm{d}\theta_{ij}\mathrm{d}\varphi_{ij} \mathrm{d}p_{\tilde{r}_{ij}} \mathrm{d}p_{\theta_{ij}} \mathrm{d}p_{\varphi_{ij}}) \nonumber \\ &\quad& \times \rho ( {\bf r}^{N_{\rm m}}, {\bf p}^{N_{\rm m}}, {\bf R}^{N_{\rm d}}, {\bf P}^{N_{\rm d}}, \tilde{r}^{N_{\rm d}}, \theta^{N_{\rm d}}, \varphi^{N_{\rm d}}, p_{\tilde{r}}^{N_{\rm d}}, p_{\theta}^{N_{\rm d}}, p_{\varphi}^{N_{\rm d}} ) (\cdots).
\end{eqnarray}
We note again that up to this point, we have just discussed an exact variable transformation. We did not modify the system itself at all.

\subsection{Bonded systems}
\label{sec:bonding}

Using the notations introduced above, we now consider bonded systems. We freeze $\tilde{r}^{N_{\rm d}}$ from an equilibrium configuration, i.e.,
$\tilde{r}^{N_{\rm d}}$ plays the same role as the positions of the pinned particles in the random pinning approach.
Thus we now can discuss the statistical mechanics of the remaining degrees of freedom, namely a mixture of monomers and dimers for a particular realization of $\tilde{r}^{N_{\rm d}}$.
The Hamiltonian of this setting, $H'$, is given by
\begin{eqnarray}
H'( {\bf r}^{N_{\rm m}}, {\bf p}^{N_{\rm m}}, {\bf R}^{N_{\rm d}}, {\bf P}^{N_{\rm d}}, \tilde{r}^{N_{\rm d}}, \theta^{N_{\rm d}}, \varphi^{N_{\rm d}}, p_{\theta}^{N_{\rm d}}, p_{\varphi}^{N_{\rm d}} ) = \qquad \qquad \qquad \qquad \qquad \qquad \qquad \nonumber \\
\sum_{i \in \mathcal{M}} \frac{{\bf p}_i^2}{2 m_i} + \sum_{\substack{i,j \in \mathcal{D} \\ (i<j)}} \left( \frac{{\bf P}_{ij}^2}{2 M_{ij}} + \frac{p_{\theta_{ij}}^2}{2I_{ij}} + \frac{p_{\varphi_{ij}}^2}{2 I_{ij}\sin^2\theta_{ij}} \right) + U( {\bf r}^{N_{\rm m}}, {\bf R}^{N_{\rm d}}, \tilde{r}^{N_{\rm d}}, \theta^{N_{\rm d}}, \varphi^{N_{\rm d}} ) \nonumber \\
= H( {\bf r}^{N_{\rm m}}, {\bf p}^{N_{\rm m}}, {\bf R}^{N_{\rm d}}, {\bf P}^{N_{\rm d}}, \tilde{r}^{N_{\rm d}}, \theta^{N_{\rm d}}, \varphi^{N_{\rm d}}, p_{\tilde{r}}^{N_{\rm d}}, p_{\theta}^{N_{\rm d}}, p_{\varphi}^{N_{\rm d}} )-\sum_{\substack{i,j \in \mathcal{D} \\ (i<j)}} \frac{p_{\tilde{r}_{ij}}^2}{2 \mu_{ij}}.
\label{eq:H'}
\end{eqnarray}
The partition function $Z'$ for a given realization $\tilde{r}^{N_{\rm d}}$,  is
\begin{eqnarray}
Z'(\tilde{r}^{N_{\rm d}}) = \int(\prod_{i \in \mathcal{M}} \mathrm{d}{\bf r}_i \mathrm{d}{\bf p}_i) \int (\prod_{\substack{i,j \in \mathcal{D} \\ (i<j)}} \mathrm{d} {\bf R}_{ij}\mathrm{d} {\bf P}_{ij}) \int (\prod_{\substack{i,j \in \mathcal{D} \\ (i<j)}} \mathrm{d}\theta_{ij}\mathrm{d}\varphi_{ij} \mathrm{d}p_{\theta_{ij}} \mathrm{d}p_{\varphi_{ij}}) \nonumber \\ \times \exp[-\beta H'( {\bf r}^{N_{\rm m}}, {\bf p}^{N_{\rm m}}, {\bf R}^{N_{\rm d}}, {\bf P}^{N_{\rm d}}, \tilde{r}^{N_{\rm d}}, \theta^{N_{\rm d}}, \varphi^{N_{\rm d}}, p_{\theta}^{N_{\rm d}}, p_{\varphi}^{N_{\rm d}} )].
\label{eq:Z'}
\end{eqnarray}
The corresponding conditional probability distribution given $\tilde{r}^{N_{\rm d}}$ is written by
\begin{eqnarray}
\rho ( {\bf r}^{N_{\rm m}}, {\bf p}^{N_{\rm m}}, {\bf R}^{N_{\rm d}}, {\bf P}^{N_{\rm d}}, \theta^{N_{\rm d}}, \varphi^{N_{\rm d}}, p_{\theta}^{N_{\rm d}}, p_{\varphi}^{N_{\rm d}} \ | \ \tilde{r}^{N_{\rm d}}) = \qquad \qquad \qquad \qquad \nonumber \\
 \frac{1}{Z'(\tilde{r}^{N_{\rm d}})} \exp[-\beta H'( {\bf r}^{N_{\rm m}}, {\bf p}^{N_{\rm m}}, {\bf R}^{N_{\rm d}}, {\bf P}^{N_{\rm d}}, \tilde{r}^{N_{\rm d}}, \theta^{N_{\rm d}}, \varphi^{N_{\rm d}}, p_{\theta}^{N_{\rm d}}, p_{\varphi}^{N_{\rm d}} )].
 \label{eq:prob_cond}
\end{eqnarray}
Hence a thermal average for a particular realization $\tilde{r}^{N_{\rm d}}$ is defined by
\begin{eqnarray}
\langle \cdots \rangle_{\tilde{r}^{N_{\rm d}}} &=& \int(\prod_{i \in \mathcal{M}} \mathrm{d}{\bf r}_i \mathrm{d}{\bf p}_i) \int (\prod_{\substack{i,j \in \mathcal{D} \\ (i<j)}} \mathrm{d} {\bf R}_{ij}\mathrm{d} {\bf P}_{ij}) \int (\prod_{\substack{i,j \in \mathcal{D} \\ (i<j)}} \mathrm{d}\theta_{ij}\mathrm{d}\varphi_{ij} \mathrm{d}p_{\theta_{ij}} \mathrm{d}p_{\varphi_{ij}}) \nonumber \\ &\quad& \times \rho ( {\bf r}^{N_{\rm m}}, {\bf p}^{N_{\rm m}}, {\bf R}^{N_{\rm d}}, {\bf P}^{N_{\rm d}}, \theta^{N_{\rm d}}, \varphi^{N_{\rm d}}, p_{\theta}^{N_{\rm d}}, p_{\varphi}^{N_{\rm d}} \ | \ \tilde{r}^{N_{\rm d}}) (\cdots).
\label{eq:thermal_average}
\end{eqnarray}
Because the choice of the bonds $\tilde{r}^{N_{\rm d}}$ that are frozen depends on the other degrees of freedom (i.e., only neighboring particles are chosen to be bonded), we cannot use a simple probability chain rule to relate the thermal average at fixed $\tilde{r}^{N_{\rm d}}$ with the total thermal average as done in random pinning~\cite{scheidler2004relaxation}. We will discuss this point in more detail in a future publication.

\section{Simulation methods}

\subsection{Model}

We employ the Kob-Andersen binary mixture~\cite{kob1995testing}, in which particles interact through the Lennard-Jones potential,
\begin{equation}\label{eqn:lj}
u_{\alpha\beta}(r) = 4\epsilon_{\alpha\beta}\left[
  {\left( \frac{\sigma_{\alpha\beta}}{r} \right)}^{12} - {\left(
    \frac{\sigma_{\alpha\beta}}{r} \right)}^6 \right],
\end{equation}
where $\alpha, \beta = A,B$ are species indexes.
Both species have the same mass, which is set to $m=1$.
The value of the parameters $\sigma_{\alpha\beta}$ and $\epsilon_{\alpha\beta}$ are given in Ref.~\cite{kob1995testing}.
The units of length and energy are set by the parameters $\sigma=\sigma_{\rm AA}=1$ and $\epsilon=\epsilon_{\rm AA}=1$, respectively, and we set the Boltzmann constant $k_{\rm B}=1$.
The potentials are cut and shifted at a distance $2.5\sigma_{\alpha\beta}$.
We simulate systems composed of $N$ particles in a cubic box of side $L$ with periodic boundary conditions at a number density $\rho=N/V=1.2$.
We use the system size $N=1200$ for the study of equilibrium dynamics and $N=32400$ for the non-equilibrium heating-cooling process and athermal mechanical test.
We use the LAMMPS simulation package~\cite{plimpton2007lammps} to generate equilibrium configurations of the original KA model.

\subsection{Making randomly bonded systems}

\begin{figure}[htbp]
\includegraphics[width=0.4\columnwidth]{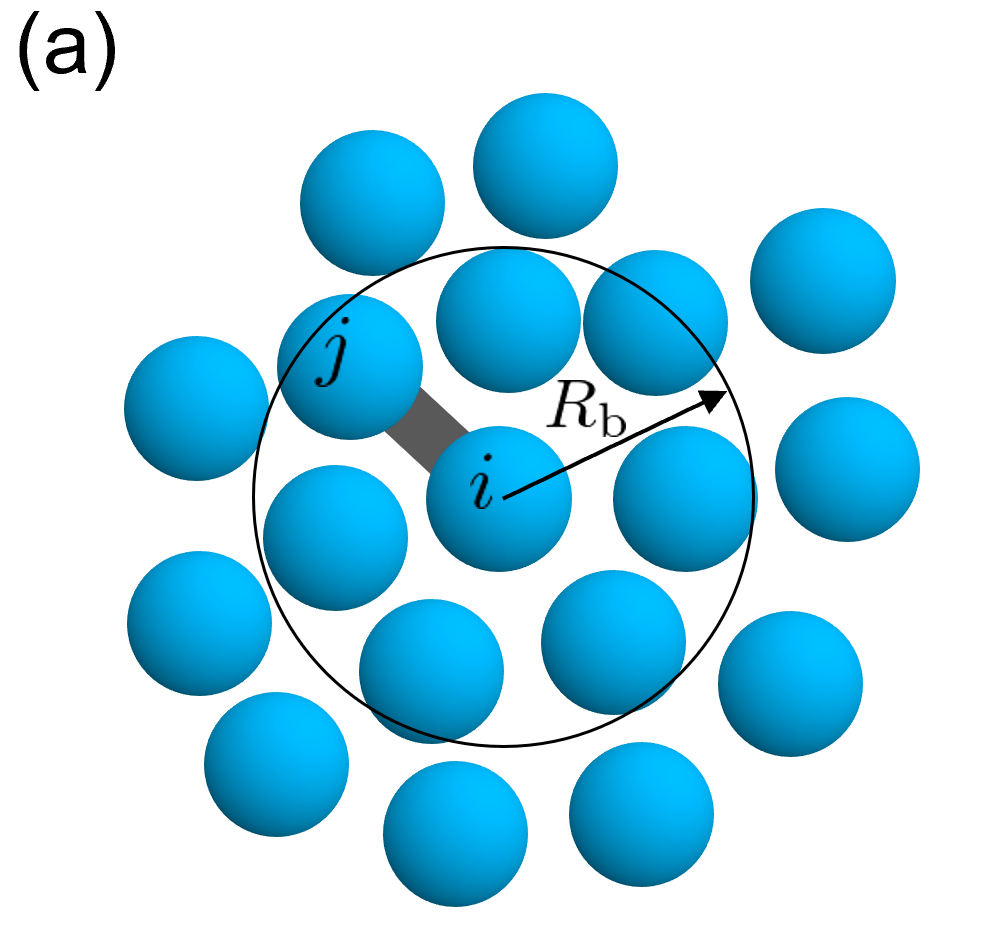}
\qquad
\includegraphics[width=0.4\columnwidth]{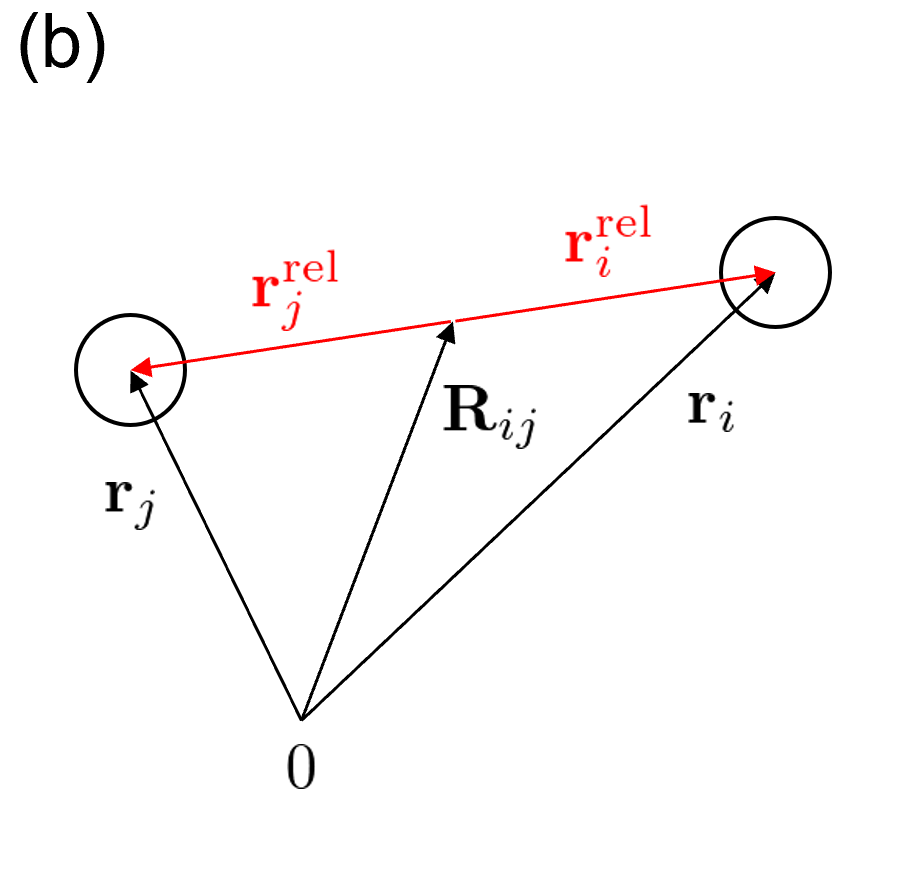}
\caption{
(a): Schematic illustration of the construction of a bond connecting particles $i$ and $j$. The sphere at the center of the particle $i$ with the radius $R_{\rm b}$ defines its neighborhood.
(b): The coordinates relative to the center of mass.
}
\label{fig:bonding_schematic}
\end{figure}

Starting from an equilibrium configuration of the original KA model with $N$ particles described above, we generate a randomly bonded system composed of monomers and dimers.
The algorithm to do so is as follows. 
First, we choose a particle randomly, say particle $i$. We then choose another particle $j$ randomly among the neighboring particles of particle $i$, 
located inside a sphere with the radius $R_{\rm b}$ and which is not yet bonded,
as schematically shown in Fig.~\ref{fig:bonding_schematic}a. 
We set $R_{\rm b}=1.5$, which is near the first minimum of the radial distribution function, thus corresponding roughly to the boundary of the first coordination shell.
We then freeze the distance between the two particles, $\tilde r_{ij} = |{\bf r}_i-{\bf r}_j|$, permanently, which means that the particles $i$ and $j$ now form a dimer. 
We repeat the above process for the remaining monomer particles until the number of dimers, $N_{\rm d}$, reaches the target value.
By construction, we have $N=N_{\rm m} + 2N_{\rm d}$, where $N_{\rm m}$ is the number of monomers. We introduce the control parameter $c=\frac{2N_{\rm d}}{N}=\frac{N-N_{\rm m}}{N}$, such that
$c=0$ corresponds to the system with only monomers (hence the original non-bonded system), whereas $c=1$ corresponds to a system having only dimers.
Using the algorithm explained above, it is difficult in practice to reach $c=1$, because at some point we run out of neighboring pairs, leaving a few percent of monomer particles. Thus we use $c=0.95$ as the maximum value that we study in this work.

\subsection{Constrained dynamics}

We implement a constrained molecular dynamics simulation for the equilibrium dynamics and non-equilibrium heating-cooling process of the bonded system via the RATTLE method~\cite{andersen1983rattle}. 
Since we are not aware of previous implementations of RATTLE with a simple Nos\'e-Hoover thermostat, we describe the details of our implementation below.

We recall the velocity-Verlet version of the Nos\'e-Hoover thermostat~\cite{frenkel2001understanding}.
\begin{eqnarray}
{\bf r}_i(t+\Delta t) &=& {\bf r}_i(t) + {\bf v}_i(t) \Delta t + \frac{(\Delta t)^2}{2} \left\{{\bf f}_i(t)-\zeta(t){\bf v}_i(t)\right\}, \\
{\bf v}_i(t+\Delta t/2) &=& {\bf v}_i(t) + \frac{\Delta t}{2}\left\{{\bf f}_i(t) - \zeta(t) {\bf v}_i(t)\right\}, \\
\zeta(t+\Delta t) &=& \zeta(t)+\frac{\Delta t}{Q} \left[ \sum_{i=1}^N \frac{1}{4}\left\{ {\bf v}_i^2(t)+{\bf v}_i^2(t+\Delta t/2) \right\} - \frac{M+1}{2} T \right], \\
{\bf v}_i(t+\Delta t) &=& \frac{1}{1+\frac{\Delta t}{2} \zeta(t+\Delta t)} \left[ {\bf v}_i(t+\Delta t/2) + \frac{\Delta t}{2} {\bf f}_i(t+\Delta t) \right],
\end{eqnarray}
where $\zeta$ is the degree of freedom of the thermostat and $M$ is the number of degrees of freedom for the kinetic part.
$M=3N$ for the standard particle systems and $M=3N_{\rm m}+5N_{\rm d}$ for bonded systems in three dimensions.
$Q$ is a parameter that plays the role of a mass for the thermostat. We set $Q=5$ and $100$ for the system with $N=1200$ and $32400$, respectively. 
We set $\Delta t=0.001$.
The updates for the position and velocity can be combined in one equation, respectively:
\begin{eqnarray}
{\bf r}_i(t+\Delta t) &=& {\bf r}_i(t) + {\bf v}_i(t) \Delta t + \frac{(\Delta t)^2}{2} \left\{ {\bf f}_i(t)-\zeta(t){\bf v}_i(t) \right\}, \\
{\bf v}_i(t+\Delta t) &=& \frac{1}{1+\frac{\Delta t}{2} \zeta(t+\Delta t)} \left[ {\bf v}_i(t) + \frac{\Delta t}{2}\left\{ {\bf f}_i(t) + {\bf f}_i(t+\Delta t) -\zeta(t) {\bf v}_i(t) \right\} \right].
\end{eqnarray}

\vspace{0.5cm}
We now impose a rigid-body constraint between particles $i$ and $j$.
The position and velocity are constrained as follows:
\begin{eqnarray}
\chi_{ij}(t) &=& \tilde{\bf r}_{ij}^2(t) - a_{ij}^2 = 0 \label{eq:constraint1}, \\
\frac{1}{2} \frac{\mathrm{d} \chi_{ij}(t)}{\mathrm{d} t} &=& \ \tilde{\bf r}_{ij} (t) \cdot \tilde{\bf v}_{ij} (t) = 0, \label{eq:constraint2}
\end{eqnarray}
where $\tilde{\bf r}_{ij} (t)={\bf r}_i(t)-{\bf r}_j(t)$, $\tilde{\bf v}_{ij} (t)={\bf v}_i(t)-{\bf v}_j(t)$, and $a_{ij}$ is the fixed bond length determined at $t=0$, namely, $a_{ij}=|{\bf r}_i(t=0)-{\bf r}_j(t=0)|$.
Below, we will also use the following notation, $\tilde{\bf f}_{ij} (t)={\bf f}_i(t)-{\bf f}_j(t)$.

We add the constraint in Eq.~(\ref{eq:constraint1}) with the associated Lagrange multiplier $\lambda_{ij}$ to the original potential $U$:
\begin{equation}
U  \to U - \frac{\lambda_{ij}}{2} \chi_{ij}.
\end{equation}
The extra term produces a force parallel to the relative position, as given by
\begin{equation}
- \nabla_i \left( - \frac{\lambda_{ij}}{2} \chi_{ij} \right) = \lambda_{ij} \tilde{\bf r}_{ij}.
\end{equation}
We also add the constraint in Eq.~(\ref{eq:constraint2}) with the associated Lagrange multiplier $\mu_{ij}$.
After that, the update equations for the position and velocity become
\begin{eqnarray}
{\bf r}_i(t+\Delta t) &=& {\bf r}_i(t) + {\bf v}_i(t) \Delta t + \frac{(\Delta t)^2}{2} \left\{ {\bf f}_i(t) + \lambda_{ij} \tilde{\bf r}_{ij}(t) - \zeta(t) {\bf v}_i(t) \right\}, \label{eq:NHR_r} \\
{\bf v}_i(t+\Delta t) &=& \frac{1}{1+\frac{\Delta t}{2}\zeta(t+\Delta t)} \left[ {\bf v}_i(t) + \frac{\Delta t}{2} \left\{ {\bf f}_i(t) + \lambda_{ij} \tilde{\bf r}_{ij}(t) + {\bf f}_i(t+\Delta t) + \mu_{ij} \tilde{\bf r}_{ij}(t+\Delta t) -\zeta(t){\bf v}_i(t) \right\} \right]. \nonumber \\
\label{eq:NHR_v}
\end{eqnarray}
The main task below is to determine $\lambda_{ij}$ and $\mu_{ij}$ so that the constraint equations in Eqs.~(\ref{eq:constraint1}) and (\ref{eq:constraint2}) always hold during the simulation.

We first determine $\lambda_{ij}$ so that $\chi_{ij}=0$ in Eq.~(\ref{eq:constraint1}) is satisfied at time $t$ and $t+\Delta t$.
With Eq.~(\ref{eq:NHR_r}), we can write $\tilde{\bf r}_{ij}(t+\Delta t)$ as
\begin{eqnarray}
\tilde{\bf r}_{ij}(t+\Delta t) &=& \tilde{\bf r}_{ij}(t) + \tilde{\bf v}_{ij}(t) \Delta t + \frac{(\Delta t)^2}{2} \left\{ \tilde{\bf f}_{ij}(t) + 2 \lambda \tilde{\bf r}_{ij}(t) - \zeta(t) \tilde{\bf v}_{ij}(t) \right\} \nonumber \\
&=& A \ \tilde{\bf r}_{ij}(t) + B \ \tilde{\bf v}_{ij}(t) + C \ \tilde{\bf f}_{ij}(t),
\end{eqnarray}
where $A = 1 + (\Delta t)^2 \lambda$, $B = \Delta t - \frac{(\Delta t)^2}{2} \zeta(t)$, and $C = \frac{(\Delta t)^2}{2}$. 
Note that we used $\lambda=\lambda_{ij}=\lambda_{ji}$ due to the Newton's third law.
Therefore, the constraint equation in Eq.~(\ref{eq:constraint1}) at $t+ \Delta t$ becomes
\begin{eqnarray}
\chi_{ij} (t+\Delta t) &=& \tilde{\bf r}_{ij}^2(t+\Delta t) - a_{ij}^2 \nonumber \\
&=& (A \ \tilde{\bf r}_{ij}(t) + B \ \tilde{\bf v}_{ij}(t) + C \ \tilde{\bf f}_{ij}(t))^2 - a_{ij}^2 \nonumber \\
&=& a_{ij}^2 A^2 + 2C(\tilde{\bf r}_{ij}(t) \cdot \tilde{\bf f}_{ij}(t)) A + (B \tilde{\bf v}_{ij}(t) + C \tilde{\bf f}_{ij}(t))^2 - a_{ij}^2=0.
\end{eqnarray}
We used $\tilde{\bf r}_{ij}^2(t)=a_{ij}^2$ and $\tilde{\bf r}_{ij}(t) \cdot \tilde{\bf v}_{ij}(t)=0$.
For diatomic dumbel systems, we can obtain $\lambda$ analytically as follows.
\begin{eqnarray}
\lambda &=& \frac{A-1}{(\Delta t)^2}, \\
A &=& \frac{-C(\tilde{\bf r}_{ij}(t) \cdot \tilde{\bf f}_{ij}(t)) + \sqrt{C^2 (\tilde{\bf r}_{ij}(t) \cdot \tilde{\bf f}_{ij}(t))^2 - a_{ij}^2(B \tilde{\bf v}_{ij}(t) + C \tilde{\bf f}_{ij}(t))^2+a_{ij}^4 }}{a_{ij}^2}.
\end{eqnarray}

Next we determine $\mu_{ij}$ so that the second constraint equation in Eq.~(\ref{eq:constraint2}) is satisfied at time $t$ and $t+\Delta t$.
With Eq.~(\ref{eq:NHR_v}), we can write $\tilde{\bf v}_{ij}(t+\Delta t)$ by
\begin{eqnarray}
\tilde{\bf v}_{ij}(t+\Delta t) &=& \frac{1}{D} \left[ \tilde{\bf v}_{ij}(t) + \frac{\Delta t}{2} \left\{ \tilde{\bf f}_{ij}(t) + 2 \lambda \tilde{\bf r}_{ij}(t) + \tilde{\bf f}_{ij}(t+\Delta t) + 2 \mu \tilde{\bf r}_{ij}(t+\Delta t) -\zeta(t)\tilde{\bf v}_{ij}(t) \right\} \right] \nonumber \\
&=& \frac{1}{D} \left[ E \ \tilde{\bf v}_{ij}(t) + \frac{\Delta t}{2} \left\{ \tilde{\bf f}_{ij}(t) + 2 \lambda \tilde{\bf r}_{ij}(t) + \tilde{\bf f}_{ij}(t+\Delta t) \right\} + \mu \Delta t \tilde{\bf r}_{ij}(t+\Delta t) \right], \nonumber
\end{eqnarray}
where $D = 1+\frac{\Delta t}{2} \zeta(t+\Delta t)$ and $E = 1 -\frac{\Delta t}{2} \zeta(t)
$.
We also used $\mu=\mu_{ij}=\mu_{ji}$.
Thus, the second constraint equation in Eq.~(\ref{eq:constraint2}) becomes
\begin{eqnarray}
\frac{1}{2} \frac{\mathrm{d} \chi_{ij}(t+\Delta t)}{\mathrm{d} t} &=& \tilde{\bf r}_{ij}(t+\Delta t) \cdot \tilde{\bf v}_{ij}(t+\Delta t) \nonumber \\
&=& \frac{1}{D} \left[ E (\tilde{\bf r}_{ij}(t+ \Delta t) \cdot \tilde{\bf v}_{ij}(t)) + \frac{\Delta t}{2} \tilde{\bf r}_{ij}(t+ \Delta t) \cdot \left\{ \tilde{\bf f}_{ij}(t) + 2 \lambda \tilde{\bf r}_{ij}(t) + \tilde{\bf f}_{ij}(t+\Delta t) \right\} + \mu \Delta t a_{ij}^2 \right] \nonumber \\
&=& 0.
\end{eqnarray}
We used $\tilde{\bf r}_{ij}^2(t+ \Delta t)=a_{ij}^2$ to derive the above equation. Finally, we obtain the expression for $\mu$ as
\begin{equation}
\mu = - \frac{1}{a_{ij}^2 \Delta t} \left[ E (\tilde{\bf r}_{ij}(t+ \Delta t) \cdot \tilde{\bf v}_{ij}(t)) + \frac{\Delta t}{2} \tilde{\bf r}_{ij}(t+ \Delta t) \cdot \left\{ \tilde{\bf f}_{ij}(t) + 2 \lambda \tilde{\bf r}_{ij}(t) + \tilde{\bf f}_{ij}(t+\Delta t) \right\} \right].
\end{equation}

\subsection{Initial velocities}

The initial configuration of a randomly bonded glass former is already in thermal equilibrium, as we confirmed numerically.
Yet, we have to assign the initial velocities so that the system follows the canonical ensemble at time $t=0$.
Under the presence of the rigid body bond constraints, we cannot use the standard expression for the Maxwell-Boltzmann distribution for each particle because it would break the constraint.
Thus we pay special attention to initializing the velocities while keeping the rigid body constraints.

We assign the velocities for the monomers, the center of mass of dimers, and the relative movements of dimers, separately, based on the Hamiltonian $H'$ in Eq.~(\ref{eq:H'}) at temperature $T=1/\beta$.
For the monomers, we can assign ${\bf v}_i(t=0)={\bf v}_i^{\rm MB}$, where ${\bf v}_i^{\rm MB}$ is drawn by a Gaussian distribution with zero mean and variance $T/m_i$.
For the center of mass of dimers, similarly, we can use ${\bf V}_{ij}(t=0) = {\bf V}_{ij}^{\rm MB}$ drawn by a Gaussian distribution with zero mean and variance $T/M_{ij}$.
Special attention has to be paid to the relative motion.
We consider the coordinates relative to the center of mass, ${\bf r}_i^{\rm rel}={\bf r}_i-{\bf R}_{ij}$ and ${\bf r}_j^{\rm rel}={\bf r}_j-{\bf R}_{ij}$, as shown schematically in Fig.~\ref{fig:bonding_schematic}b.
${\bf r}_i^{\rm rel}$ and ${\bf r}_j^{\rm rel}$ can be expressed by ${\bf r}_i^{\rm rel} = \frac{m_j}{M_{ij}} \tilde{\bf r}_{ij}$ and ${\bf r}_j^{\rm rel} = - \frac{m_i}{M_{ij}} \tilde{\bf r}_{ij}$, respectively.
Considering both the center of mass and relative motion, we assign the velocities of particles $i$ and $j$ connected by a rigid bond as follows.
\begin{eqnarray}
{\bf v}_i(t=0) &=& {\bf V}_{ij}(t=0) + \dot{\bf r}_i^{\rm rel}(t=0) = {\bf V}_{ij}^{\rm MB} + \frac{m_j}{M_{ij}} \dot{\tilde{{\bf r}}}_{ij}(t=0), \\
{\bf v}_j(t=0) &=& {\bf V}_{ij}(t=0) + \dot{\bf r}_j^{\rm rel}(t=0) = {\bf V}_{ij}^{\rm MB} - \frac{m_i}{M_{ij}} \dot{\tilde{{\bf r}}}_{ij}(t=0).
\end{eqnarray}
The remaining task is to assign $\dot{\tilde{{\bf r}}}_{ij}(t=0)$ in the above equations.
Here we drop the indexes ${ij}$ for simplicity.
We express $\tilde{\bf r}$ by the spherical coordinate in $d=3$:
\[
\tilde{\bf r} = \left\{ \begin{array}{ll}
\tilde{x} = \tilde{r} \sin \theta \cos \varphi \\
\tilde{y} = \tilde{r} \sin \theta \sin \varphi \\
\tilde{z} = \tilde{r} \cos \theta
\end{array} \right.
\]
Since $\tilde{r}$ is constant, the relative velocity $\dot{\tilde{\bf r}}$ is given as follows.
\[
\dot{\tilde{\bf r}} = \left\{ \begin{array}{ll}
\dot{\tilde{x}} = \tilde{r} (\cos \theta \cos \varphi \ \dot{\theta} - \sin \theta \sin \varphi \ \dot{\varphi}) \\
\dot{\tilde{y}} = \tilde{r} (\cos \theta \sin \varphi \ \dot{\theta} + \sin \theta \cos \varphi \ \dot{\varphi}) \\
\dot{\tilde{z}} = - \tilde{r} \sin \theta \ \dot{\theta}
\label{eq:rel_vel}
\end{array} \right.
\]
We thus initialize the values of $\dot{\theta}$ and $\dot{\varphi}$ to initialize $\dot{\tilde{{\bf r}}}$.
Because the corresponding kinetic terms in Eq.~(\ref{eq:H'}) can be written by
\begin{equation}
-\beta \left( \frac{p_{\theta}^2}{2I} + \frac{p_{\varphi}^2}{2 I\sin^2\theta} \right)=-\beta \left( \frac{I}{2} \dot{\theta}^2 + \frac{I \sin^2 \theta}{2} \dot{\varphi}^2 \right),
\end{equation}
we obtain the initial values of $\dot{\theta}$ and $\dot{\varphi}$ by the Gaussian distribution with zero mean and variance $T/I$ and $T/(I \sin^2 \theta)$, respectively.

\subsection{Time correlation functions}

To study the equilibrium dynamics, we compute the self part of the intermediate scattering functions for monomers, the center of mass of dimers, and all particles, given by
\begin{eqnarray}
F_{\rm s}^{\rm Mono}(q,t) &=& \overline{ \left\langle \frac{1}{N_{\rm m}} \sum_{i \in \mathcal{M}} e^{-i {\bf q}\cdot ({\bf r}_i(t)-{\bf r}_i(0))} \right\rangle_{\tilde{r}^{N_{\rm d}}} }, \\
F_{\rm s}^{\rm Di}(q,t) &=& \overline{ \left\langle \frac{1}{N_{\rm d}} \sum_{\substack{i,j \in \mathcal{D} \\ (i<j)}} e^{-i {\bf q}\cdot ({\bf R}_{ij}(t)-{\bf R}_{ij}(0))} \right\rangle_{\tilde{r}^{N_{\rm d}}} }, \\
F_{\rm s}^{\rm All}(q,t) &=& \overline{ \left\langle \frac{1}{N} \sum_{i=1}^N e^{-i {\bf q}\cdot ({\bf r}_i(t)-{\bf r}_i(0))} \right\rangle_{\tilde{r}^{N_{\rm d}}} },
\end{eqnarray}
respectively,
where $\overline{\cdots}$ indicates an average over different bond realizations and $\langle \cdots \rangle_{\tilde{r}^{N_{\rm d}}}$ is a time average over a given realization $\tilde{r}^{N_{\rm d}}$.
In practice we have averaged over 5–20 different realizations to calculate the time correlation functions.
We set $q=7.25$, the location of the main peak in the static structure factor~\cite{kob1995testing}.

\begin{figure}[b]
\includegraphics[width=0.48\columnwidth]{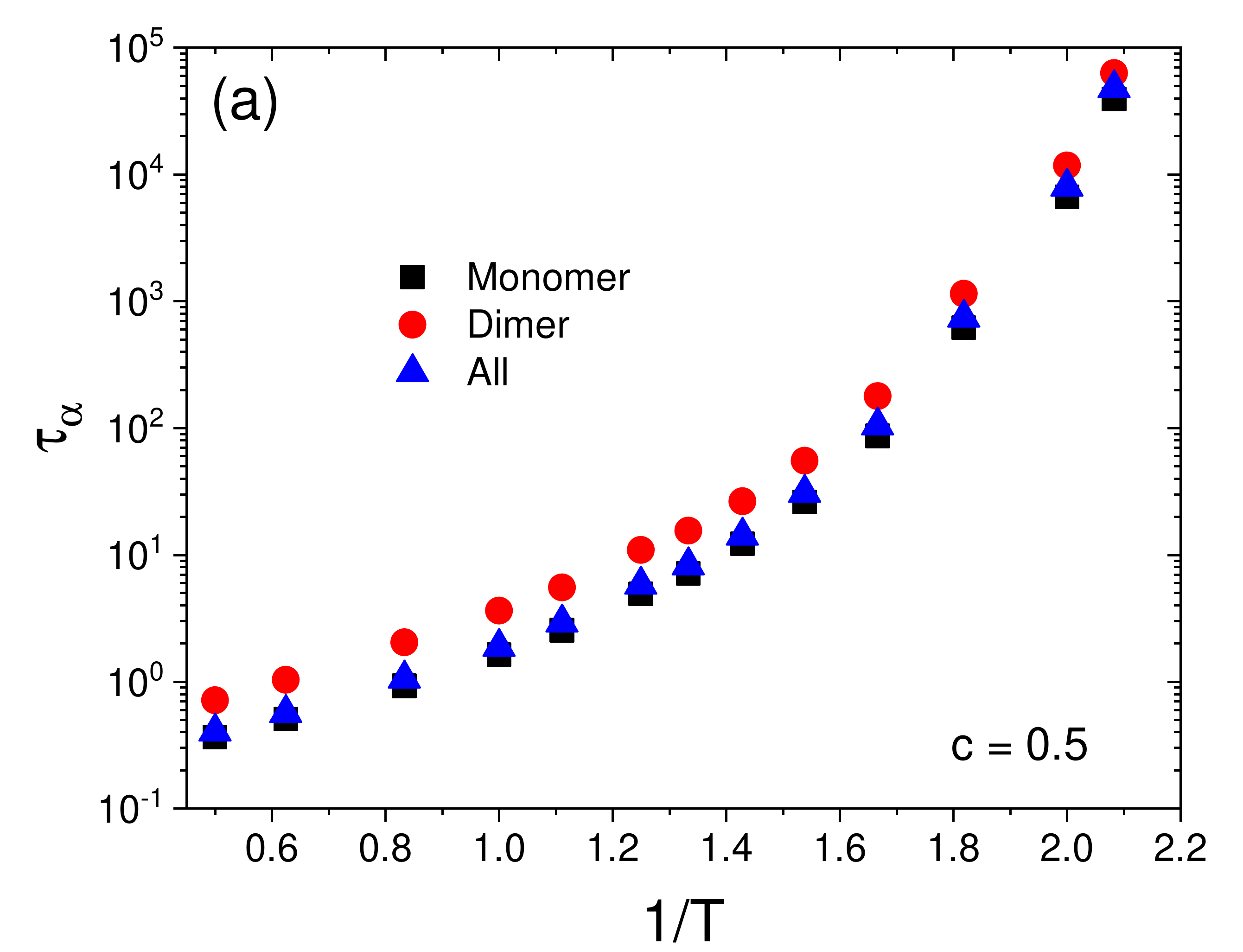}
\includegraphics[width=0.48\columnwidth]{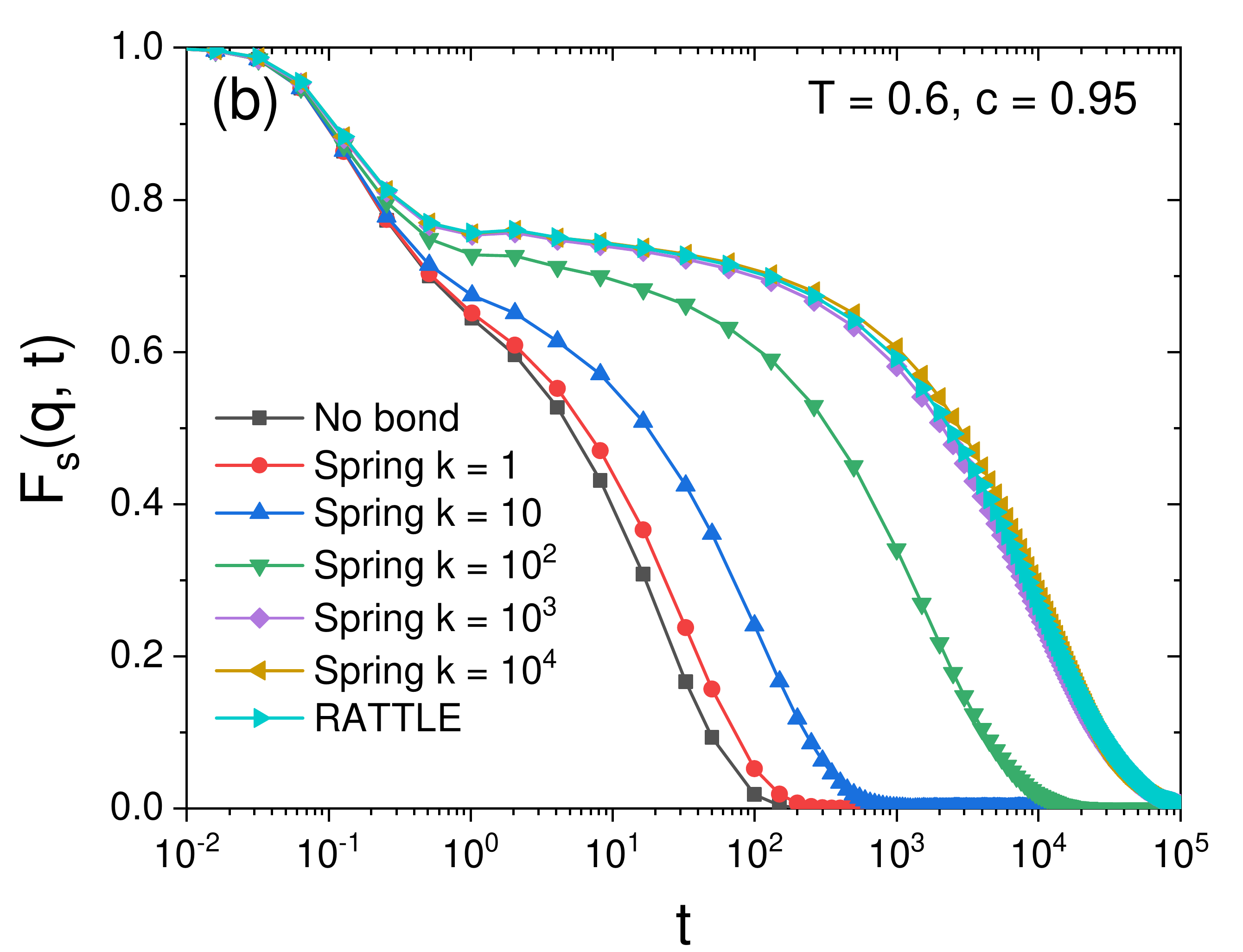}
\caption{
(a): Relaxation time $\tau_{\alpha}$ vs. the inverse of temperature $1/T$ for $c=0.5$.
(b): Simulations of harmonic spring method with several spring constant $k$. $F_{\rm s}(q,t)$ from all particles for $T=0.6$ and $c=0.95$. The system size is $N=1200$.
}
\label{fig:absence_agingSI}
\end{figure}

In the main text we have shown the intermediate scattering functions obtained from the equilibrium dynamics at $T=0.6$ and $c=0.95$, computed from the positions of monomers, $F_{\rm s}^{\rm Mono}(q,t)$, the center of mass of the dimers, $F_{\rm s}^{\rm Di}(q,t)$, and all particles, $F_{\rm s}^{\rm All}(q,t)$, together with
the corresponding mean-squared displacements and rotational correlation function (for dimer molecules).
$F_{\rm s}^{\rm Di}(q,t)$ displays a higher value of the plateau than the one found in $F_{\rm s}^{\rm Mono}(q,t)$ or $F_{\rm s}^{\rm All}(q,t)$. However, these functions relax on essentially the same timescale which is evidence that the monomers and dimers  have a quite similar relaxation dynamics. We define the relaxation time $\tau_{\alpha}$ as the time at which the intermediate scattering function decays to $1/e$ and
present the $\tau_{\alpha}$ vs. $1/T$ plot for $c=0.5$ in Fig.~\ref{fig:absence_agingSI}a.
The relaxation time $\tau_{\alpha}$ for the center of mass of dimers exceeds $\tau_\alpha$ for the monomers by a factor of two, independent of $T$, which shows that the dynamics of the two types of particles does not decouple from each other, and we have checked that this is also the case for the other values of $c$.
We conclude that the three definitions of the intermediate scattering functions provide us with essentially the same information in terms of structural relaxation. We thus decided to stick to $F_{\rm s}^{\rm All}(q,t)$ for most of our results, and drop the superscript unless otherwise specified.

In the main text, we have also presented a rotational correlation function for dimers, $C(t)$, defined by
\begin{equation}
    C(t) = \overline{ \left\langle \frac{1}{2 N_{\rm d}} \sum_{\substack{i,j \in \mathcal{D}}} {\bf n}_{ij}(t) \cdot {\bf n}_{ij}(0) \right\rangle_{\tilde{r}^{N_{\rm d}}} },
\end{equation}
where ${\bf n}_{ij}=\tilde {\bf r}_{ij}/a_{ij}$ is the unit vector of the orientation of a dumbbell molecule composed of particles $i$ and $j$.
$C(t)$ is an analog of the dynamical correlation function measured in dielectric experiments~\cite{ediger1996supercooled}.

In order to check our implementation of RATTLE, 
we perform independent molecular simulations by replacing the rigid bonds by a harmonic springs.
These harmonic springs are introduced in addition to the original potential energy, namely,
\begin{equation}
    U \to U + k \sum_{\substack{i,j \in \mathcal{D} \\ (i<j)}} (\tilde{r}_{ij}-a_{ij})^2,
\label{eq:spring}
\end{equation}
where $k$ is the spring constant. Hence
$k \to \infty$ should correspond to the rigid body constraint.
Figure~\ref{fig:absence_agingSI}b shows $F_{\rm s}(q,t)$ for $T=0.6$ and $c=0.95$ for different values of $k$ together with the results from the simulations using RATTLE.
We find that simulations with $k=10^3-10^4$ converge to the results from the simulations using RATTLE, which is evidence that our algorithm to constrain the bonds is correct.

\begin{figure}[t]
\includegraphics[width=0.32\columnwidth]{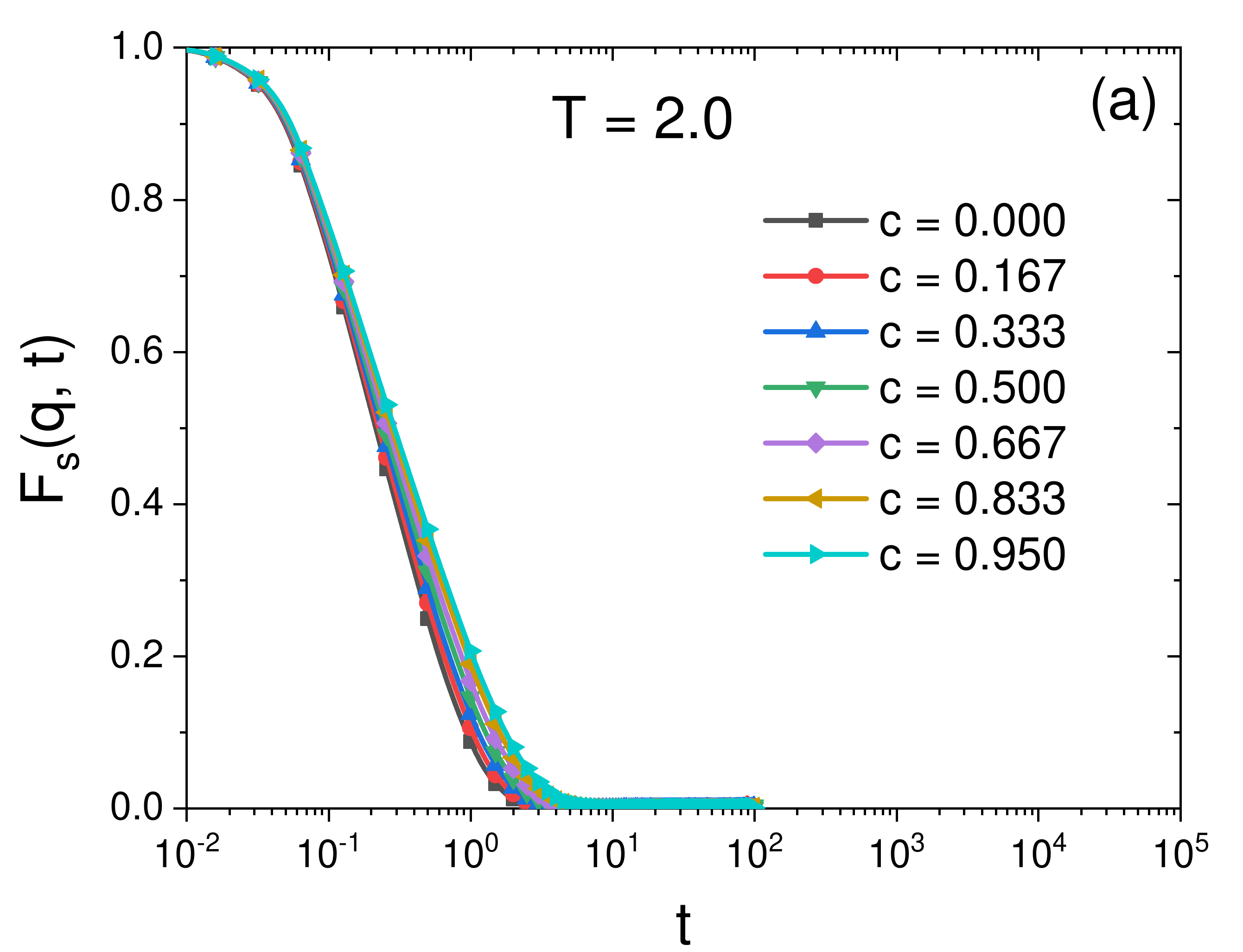}
\includegraphics[width=0.32\columnwidth]{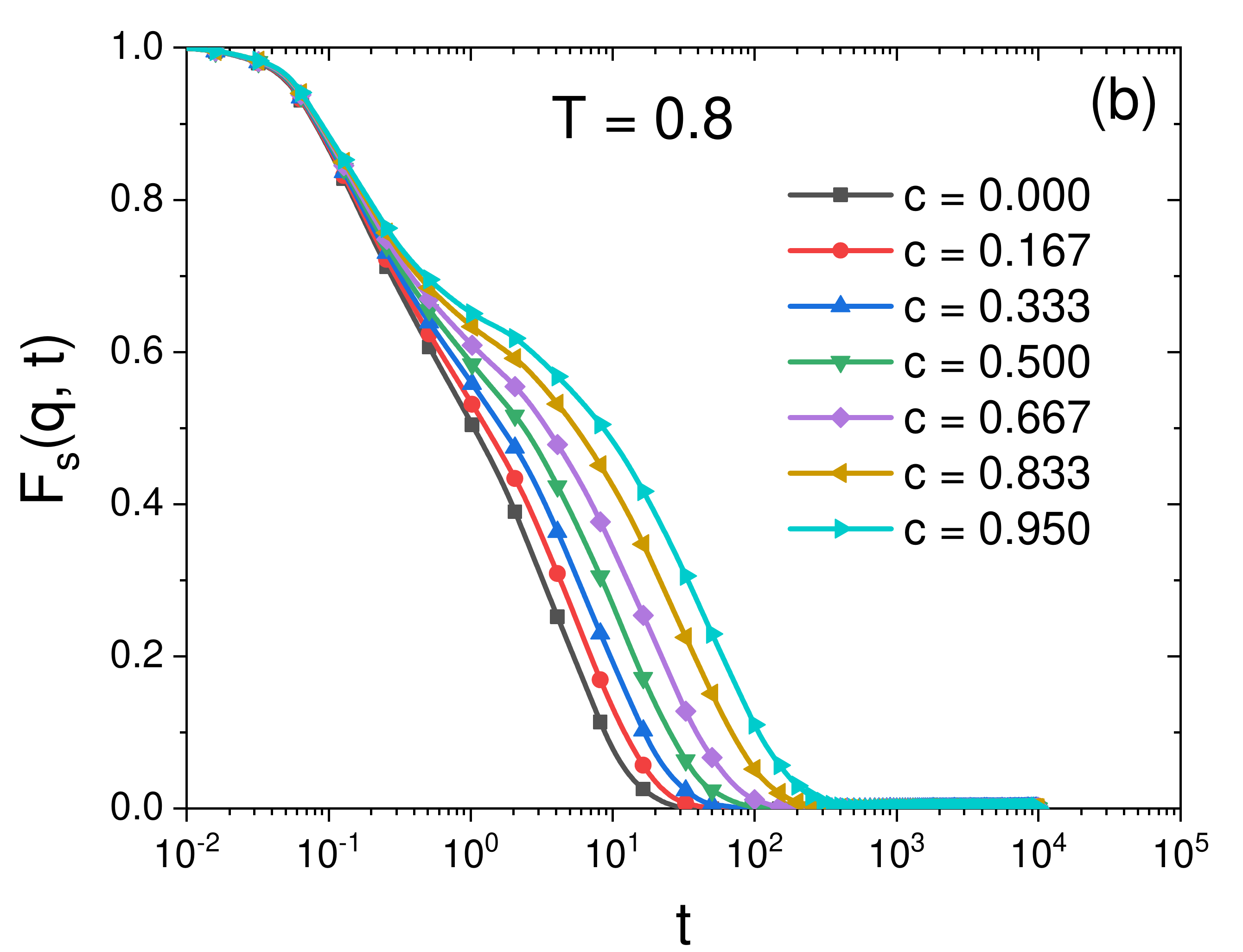}
\includegraphics[width=0.32\columnwidth]{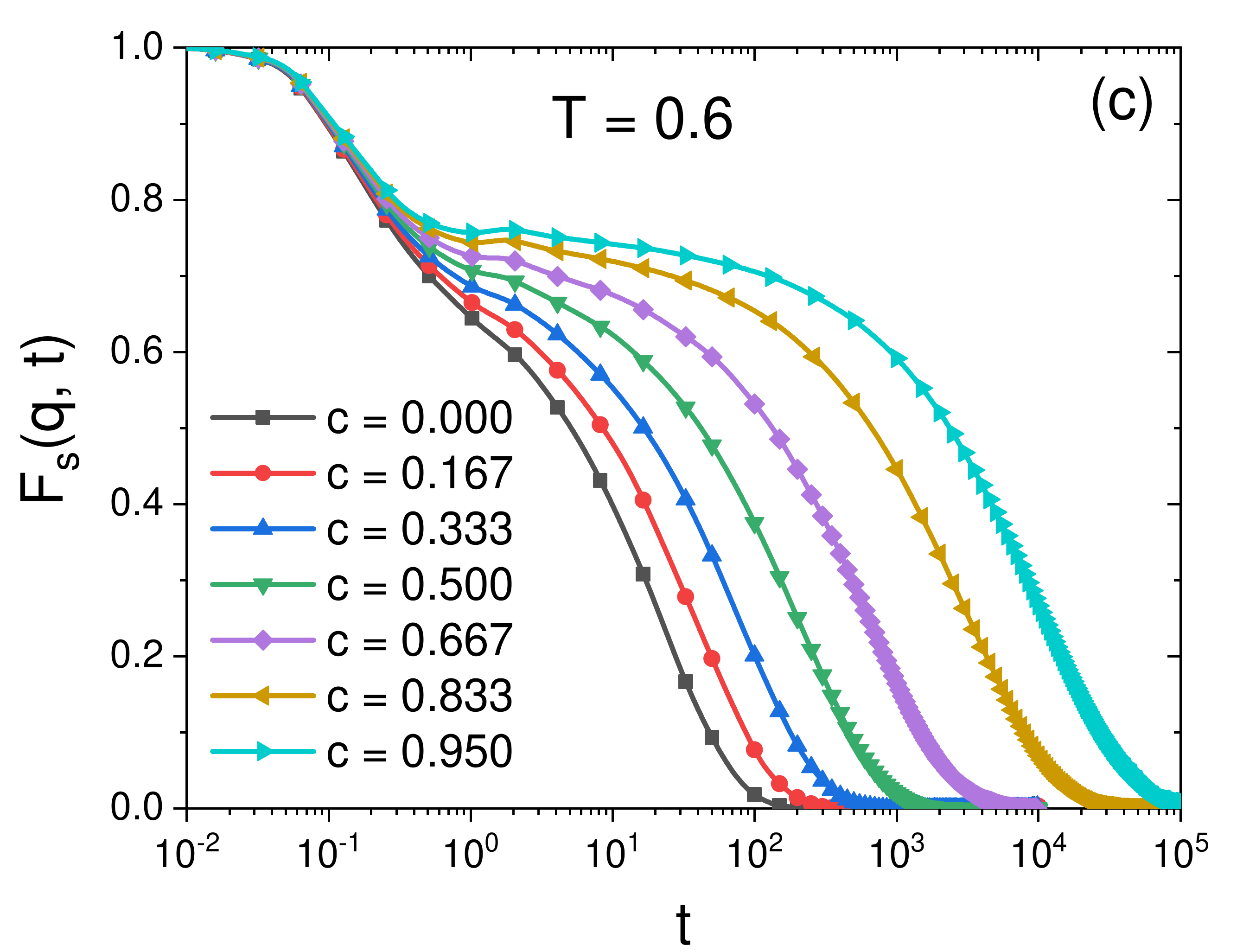}
\caption{$F_{\rm s}(q,t)$ for all particles for several $c$ for $T=2.0$ (a), $T=0.8$ (b), and $T=0.6$ (c). The system size is $N=1200$.
}
\label{fig:SI_EQ_dynamics}
\end{figure}

We present $F_{\rm s}(q,t)$ for different $T$ and $c$ in Fig.~\ref{fig:SI_EQ_dynamics} to complement the results shown in the main text.
We find that the effect of bonding is very small if temperature is high, $T=2.0$ in Fig.~\ref{fig:SI_EQ_dynamics}a.
Yet the effect is enhanced significantly as $T$ is decreased, akin to the results for randomly pinned systems~\cite{jack2013dynamical,chakrabarty2016understanding}. 
In order to see the influence of $T$ and $c$ together, we have shown the iso-$\tau_{\alpha}$ curves in the $T$ versus $c$ plane in the main text.
The iso-$\tau_{\alpha}$ curves increase with increasing $c$, which is again similar to the results found for randomly pinned systems~\cite{kob2013probing,ozawa2015equilibrium}. 

\begin{figure}[t]
\includegraphics[width=0.48\columnwidth]{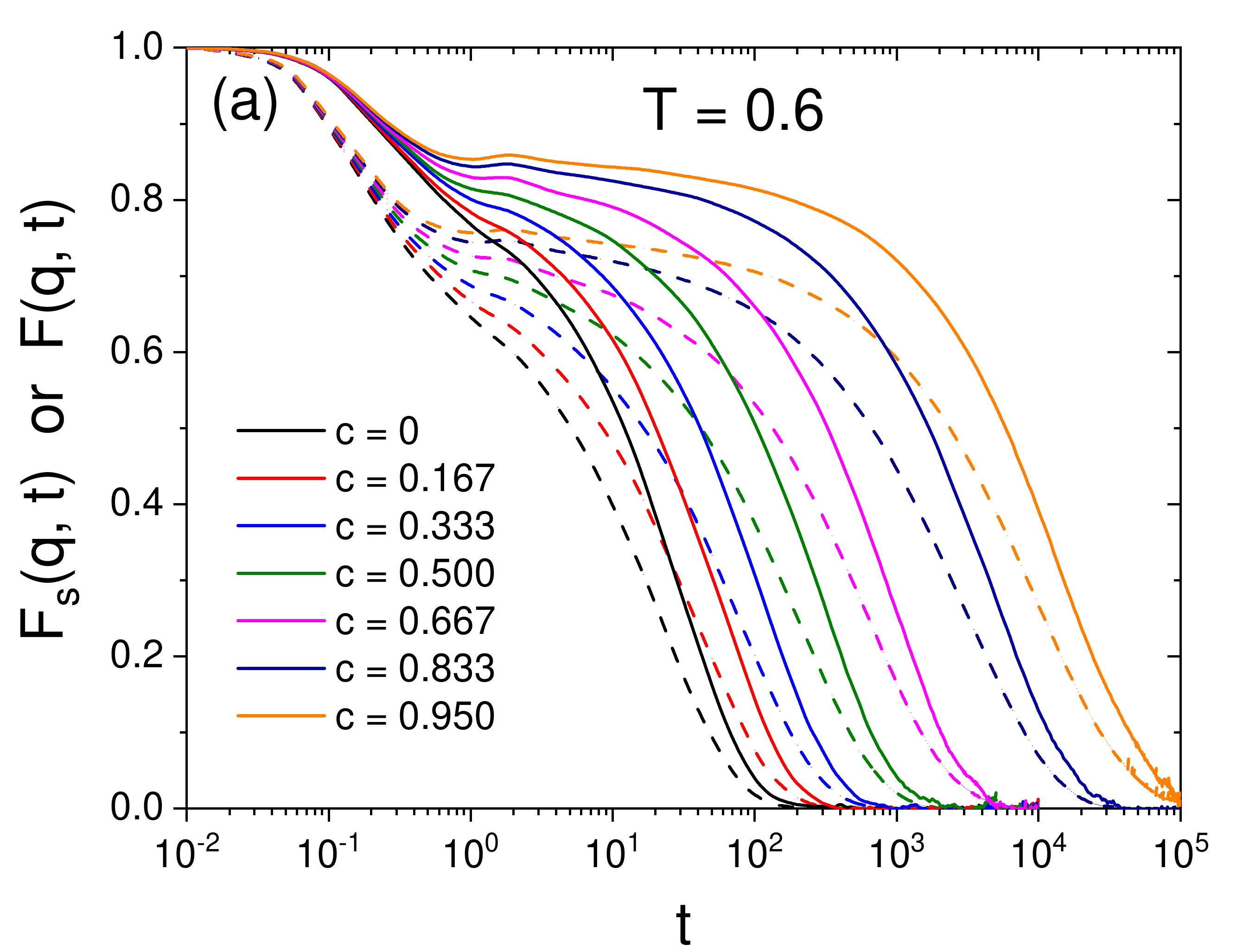}
\includegraphics[width=0.48\columnwidth]{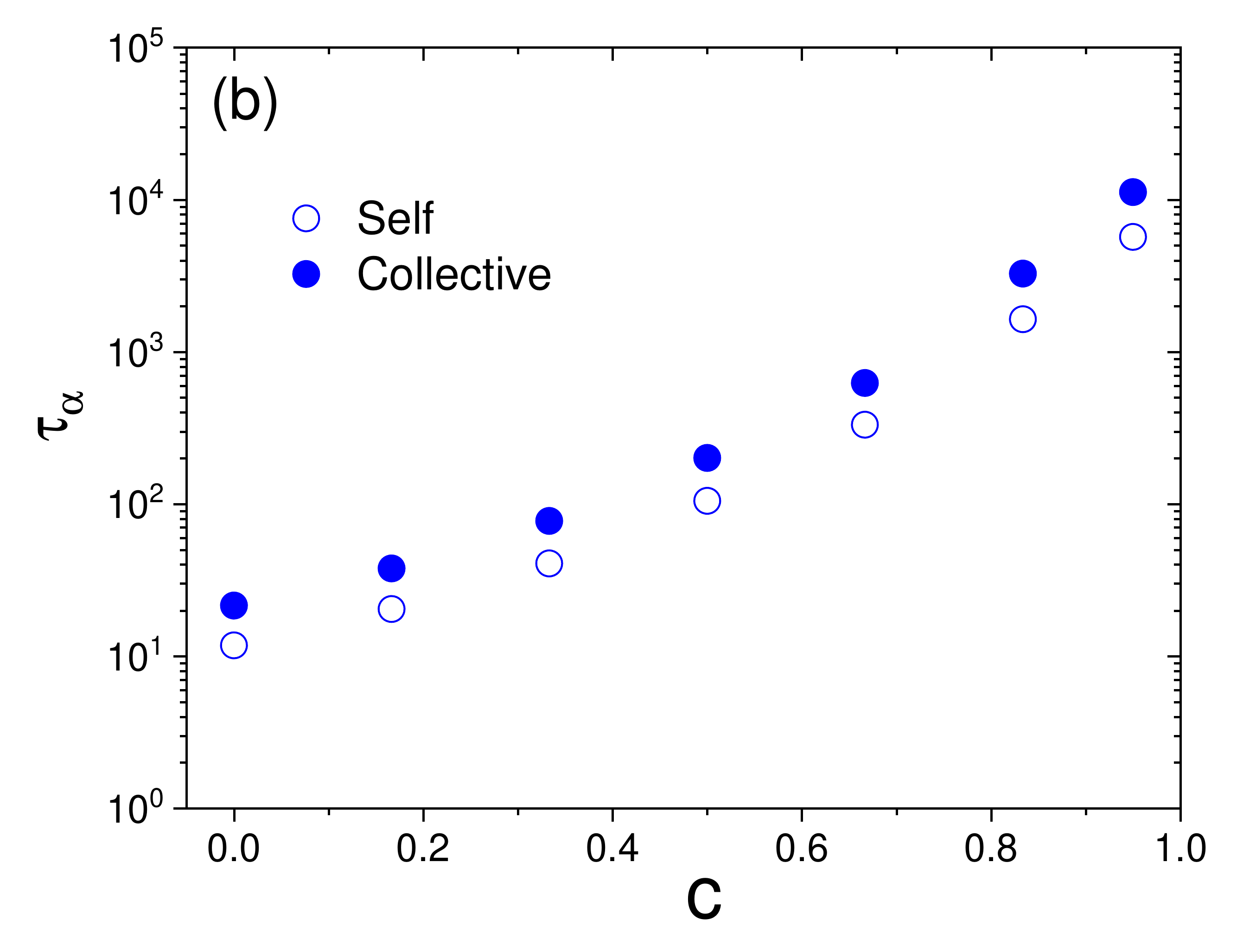}
\caption{
(a): Self (dashed curves) and collective (solid curves) intermediate scattering functions at $T=0.6$ for different $c$.
(b): Relaxation time $\tau_{\alpha}$ vs. $c$, computed from the data in (a). The system size is $N=1200$.
}
\label{fig:self_vs_collective}
\end{figure}

\begin{figure}[t]
\includegraphics[width=0.48\columnwidth]{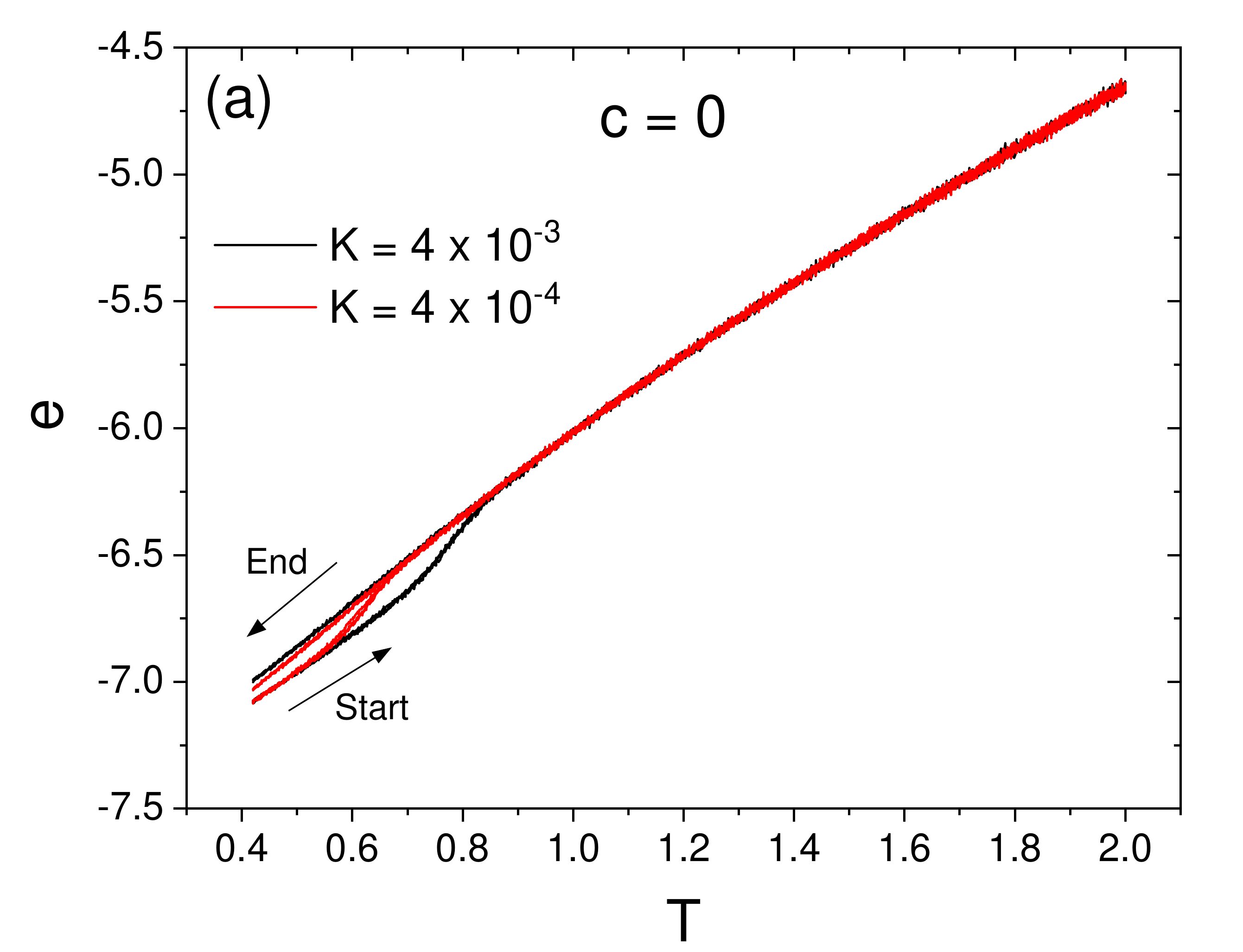}
\includegraphics[width=0.48\columnwidth]{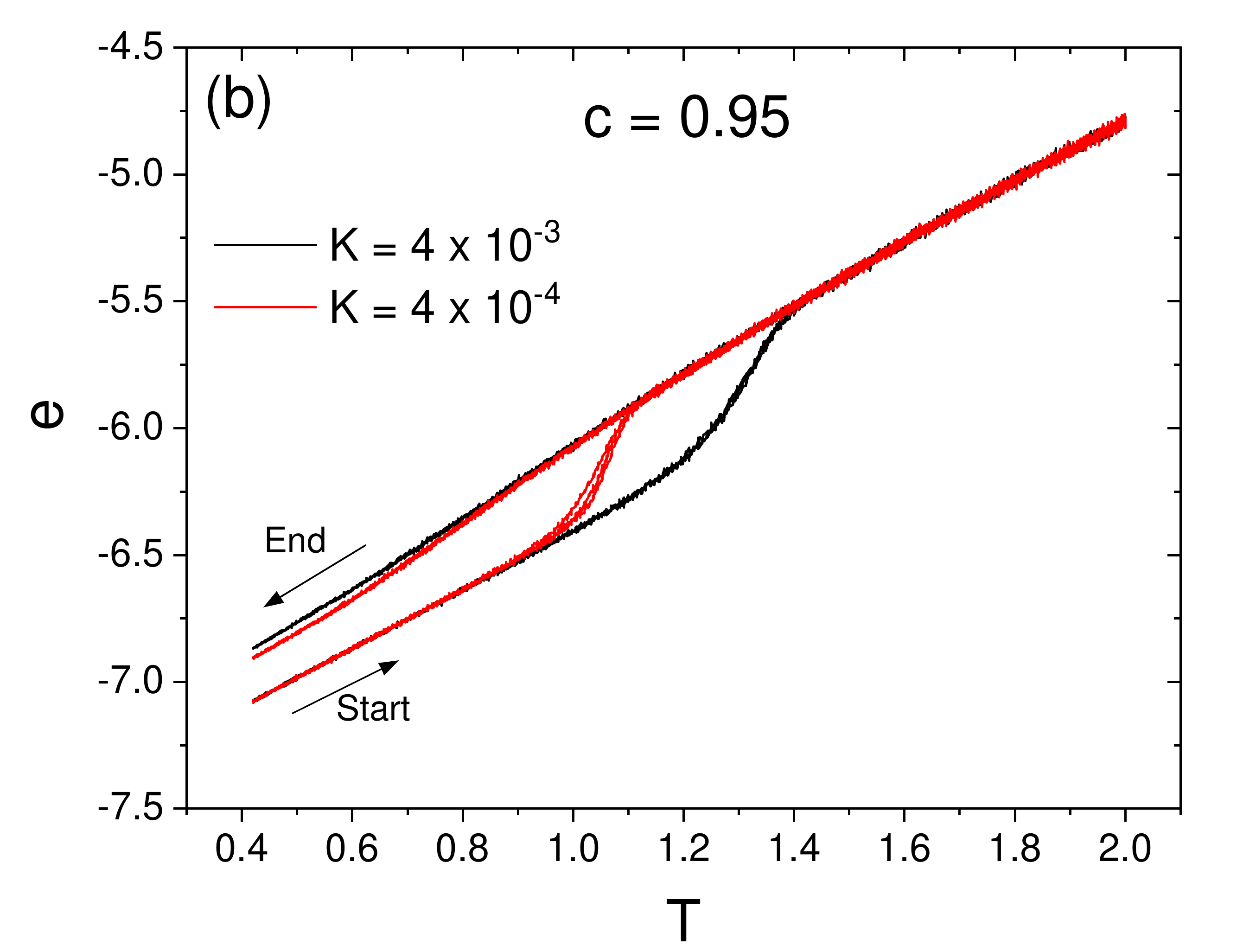}
\caption{
Heating-cooling process. The potential energy per particle $e$ as a function of temperature $T$ is shown. (a): The original system without bond ($c=0$). (b): The bonded system ($c=0.95$).
Three independent realizations are presented.
The system size is $N=32400$.
}
\label{fig:stabilitySI}
\end{figure}

\begin{figure}[t]
\includegraphics[width=0.48\columnwidth]{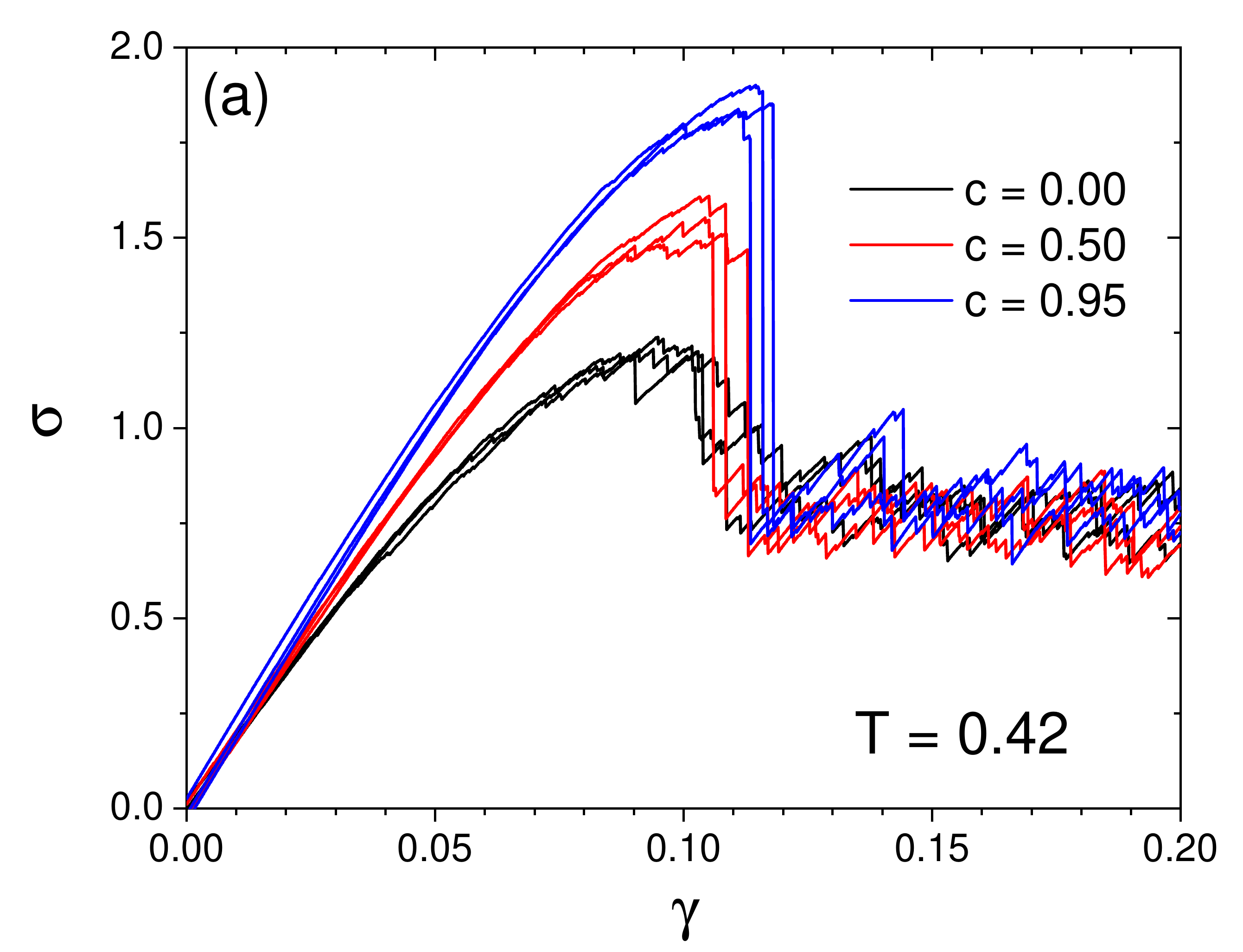}
\includegraphics[width=0.48\columnwidth]{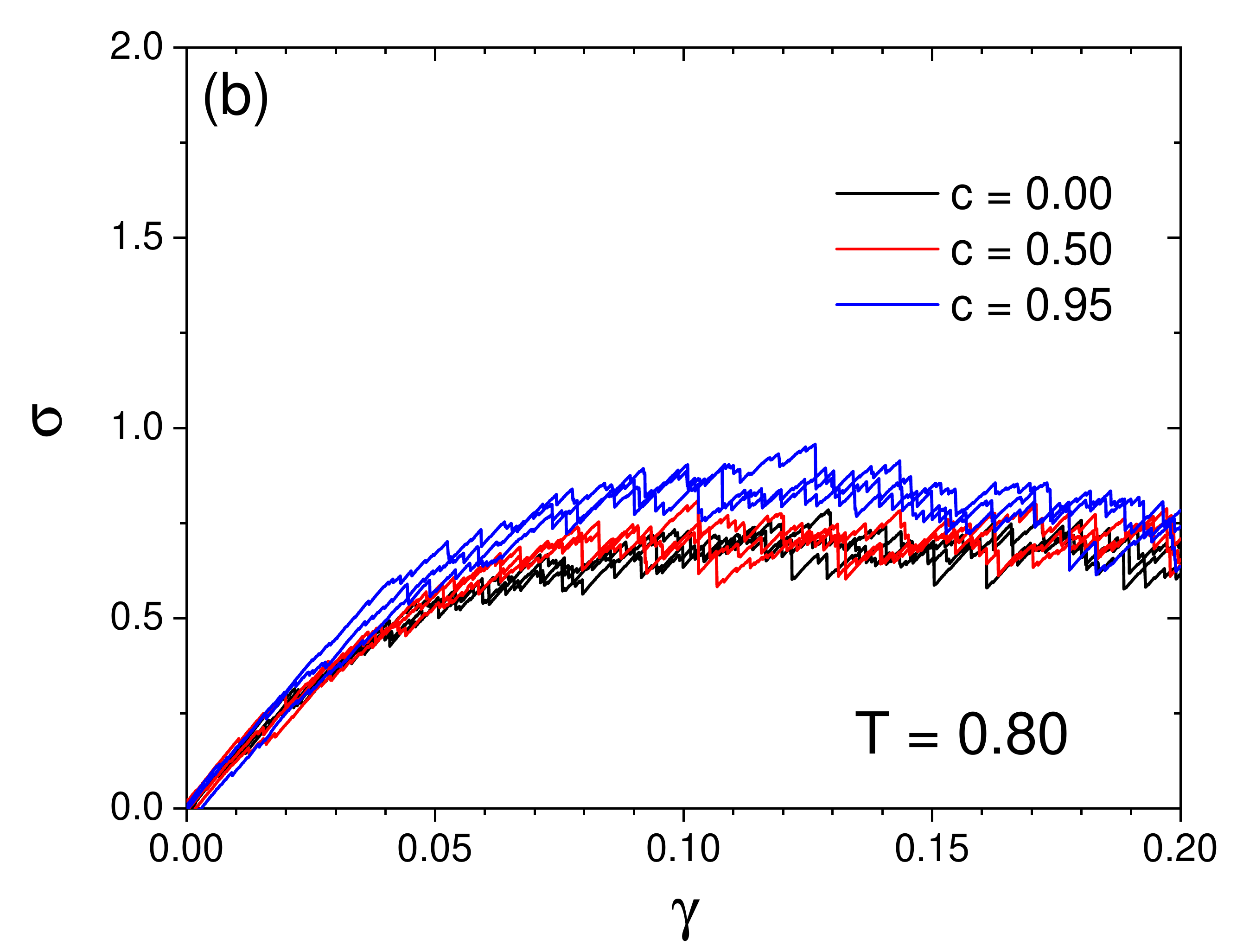}
\caption{
Shear stress $\sigma$ vs. strain $\gamma$ curves for randomly bonded glass formers generated for $c=0$, $0.5$, and $0.95$ at the preparation temperature $T=0.42$ and $T=0.8$, in panel (a) and (b), respectively.
Three individual realizations are shown.
The system size is $N=32400$.
}
\label{fig:rheology_SI}
\end{figure}

Finally we present a comparison between the self and collective parts of the intermediate scattering functions. 
It has been reported that the collective part shows apparent freezing in randomly pinned fluids, which makes the analysis difficult~\cite{ozawa2015equilibrium}. 
In Fig.~\ref{fig:self_vs_collective}a we show the self and collective parts for the randomly-bonded glass formers at $T=0.6$ for different values of $c$.
One recognizes that, in contrast to the pinned systems, the collective part also relaxes to zero and that the relaxation time is slightly larger than the one for the self part. 
Interestingly, however, the ratio between the two timescales is about a factor two irrespectively of $c$, as shown in Fig.~\ref{fig:self_vs_collective}b, i.e., self and collective correlators do not decouple. Therefore we can conclude that the self part gives reliable dynamical information about the system.

\subsection{Kinetic and mechanical stability}

In Fig.~\ref{fig:stabilitySI} we show additional supporting data for the heating/cooling process for the original (non-bonded) system ($c=0$) (a) and bonded system ($c=0.95$) (b), with two distinct values of the heating/cooling rate $K$. In both cases, we start from the system prepared at $T=0.42$, and we heat it up to $T=2.0$ (the first heating in the main text) and cool it back down to $T=0.42$. We do not report the second heating here.
The data for the non-bonded system of monomers ($c=0$, Fig.~\ref{fig:stabilitySI}a) do not show a marked hysteresis, as expected for an ordinary glass.
On the contrary, the data for the bonded system ($c=0.95$, Fig.~\ref{fig:stabilitySI}b) show a marked and strongly $K$-dependent hysteresis, characteristic of an ultrastable glass.

We also report additional data for our mechanical tests.
We firstly prepare an equilibrium configuration of a randomly bonded glass at temperature $T$ and concentration $c$.
We then perform a rapid quench down to zero temperature by using the conjugate gradient method~\cite{wright1999numerical}.
To this end, we replace the rigid bond constraint with the harmonic spring in Eq.~(\ref{eq:spring}) with sufficiently large $k$.
We vary $k$ from $10^1$ to $10^4$ and find that simulations above $k=10^3$ converge, which is consistent with the equilibrium dynamics in Fig.~\ref{fig:absence_agingSI}b.
Once we prepared a zero-temperature glass sample, we perform an athermal quasi-static shear simulation using Lees-Edwards boundary conditions~\cite{maloney2006amorphous}, still keeping a harmonic spring constraint.
This method consists of a succession of tiny uniform shear deformations with $\Delta \gamma=10^{-4}$
followed by energy minimization via the conjugate-gradient
method.

We compute the shear stress $\sigma$ by the Irvine-Kirkwood formula~\cite{allen2017computer} using the potential $U$ in Eq.~(\ref{eq:spring}). In principle the contribution from the spring term should be strictly zero in the perfectly rigid-bond simulations. However, in practice it produces a finite value of the shear stress in our harmonic spring simulations, which we neglect in the following since this additional contribution does not change the conclusions of this paper.

We present the stress vs. strain curves for $c=0$, $0.5$, and $0.95$, prepared at $T=0.42$ and $T=0.8$ in Fig.~\ref{fig:rheology_SI}. Whereas the $T=0.42$ data show sharp brittle yielding for $c=0.5$ and $0.95$, the $T=0.8$ data do not show this sharp yielding. This means that the sharp brittle yielding for $T=0.42$ originates from the fact that the system is sitting deep inside the glassy landscape,
and does not originate from the bonding process itself.

\section{Bonding in patchy colloid experiment}

In this section, we describe the experimental method and the results obtained for bonded colloidal particles dispersed in a quasi-two-dimensional sample. 
In the experiment, we use the concentration of salt to induce an attraction between the patchy colloidal particles that is sufficiently strong to create bonds between hydrophobic patches~\cite{chen2011directed}. 

The patchy particles were produced as follows~\cite{chen2011directed,chen2011triblock,iwashita2014orientational}:
First, a 5-nm-thick chromium layer followed by a 40-nm-thick gold layer was thermally deposited onto the hemispheres of spherical silica particles (2.7 $\mu$m diameter, polydispersity $2.5\%$, HIPRESICA TS, UEXC). 
The patches were chemically etched as described in Ref.~\cite{chen2011triblock}, and their size estimated from our etching-time dependence experiment is $\theta_\mathrm{ap} \sim 50^\circ$, the opening angle of a circular patch measured from its center. 
The patches were hydrophobized by 1-octadecanethiol (90\%, Sigma-Aldrich) \cite{chen2011directed}.

\begin{figure}[b]
    \begin{center}
	\includegraphics[width=0.75\textwidth]{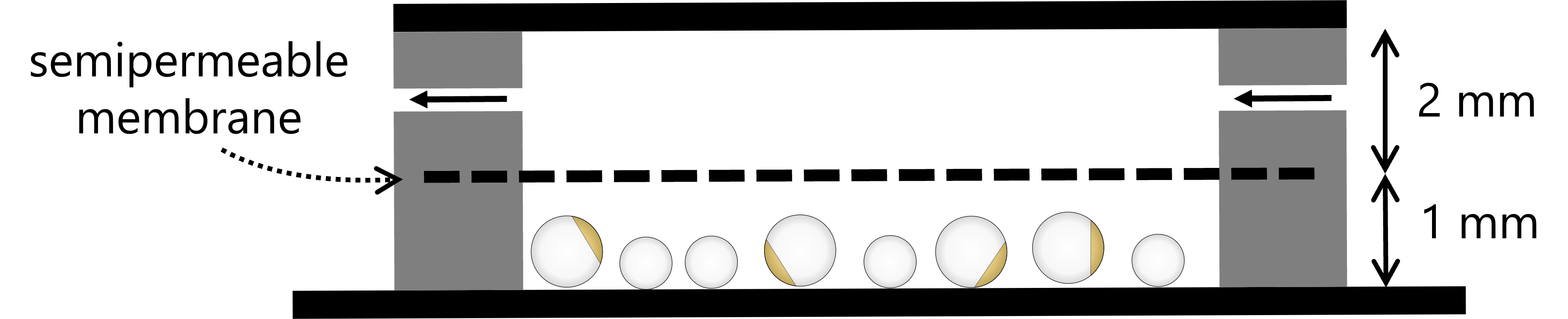}
	\caption{Sketch of the experimental cell.}
	\label{sfig:i1}
    \end{center}
\end{figure}

\begin{figure}[t]
    \begin{center}
	\includegraphics[width=0.95\textwidth]{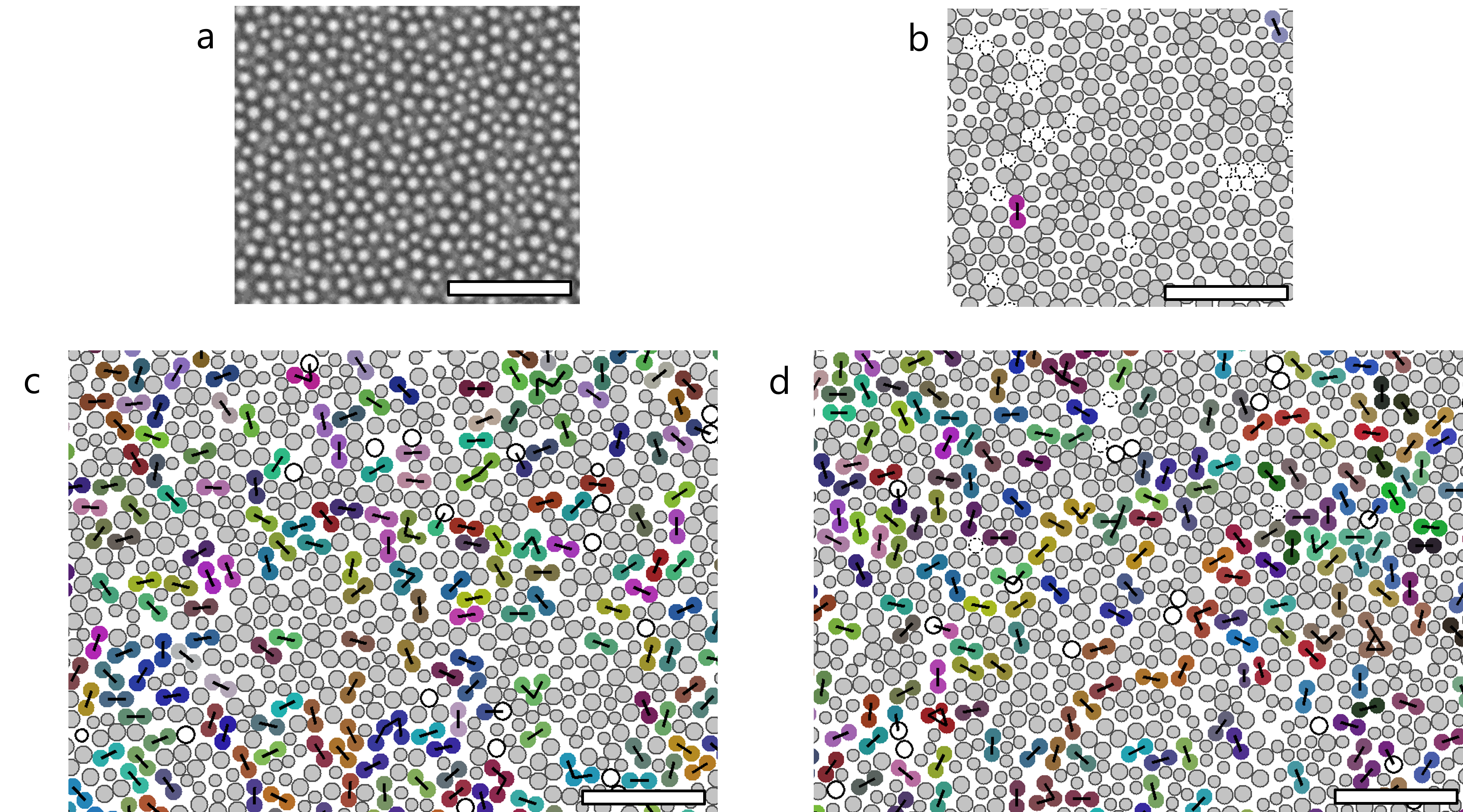}
	\caption{Two-dimensional dispersions of patchy and non-patchy particle mixture. 
	(a) A microscopy image at 0.0 mM. 
	Large particles are patchy and small particles are without patch. 
	1 \% of particles were originally bonded, survived the initial sonication.
	$\phi = 0.60$.
	(b) to (d) Reconstructed images from the analysis on particle motion. 
	The classification is described in the text.
	The salt concentration and area fraction are 0.0 mM and $\phi = 0.63$ for (b), 4.0 mM and $\phi = 0.63$ for (c) and 6.0 mM and $\phi = 0.62$ for (d).
	Scale bars: 20 $\mu$m.}
	\label{sfig:i2}
    \end{center}
\end{figure}

The produced one-patch particles were mixed with spherical silica particles (2.0 $\mu$m  $\pm 2.5\%$ in diameter, HIPRESICA TS, UEXC) in water (18.2M$\Omega$), and the particles were dispersed by ultrasonication for 1 minute. 
The dispersion of the binary particle mixture was confined in the lower layer of the experimental cell (Fig.~\ref{sfig:i1}). 
The top and bottom plates are glass slides hydrophilized by plasma cleaning. 
The upper part of the cell, which was separated from the lower part by a semipermeable membrane, had an inlet and outlet channel to exchange the liquid in the upper layer. 
Thus, by changing the NaCl concentration in the upper layer, the salt concentration in the lower layer can be tuned by the molecular exchange {\it via} the semipermeable membrane without inducing convection in the lower layer. 
We increased the NaCl concentration in the lower layer stepwise from 0.0, 1.0, 2.0, 4.0, 6.0, 8.0 to 10.0 mM, where the salt solution in the upper layer was exchanged several times in more than an hour for each concentration step. 
The particle dispersion was equilibrated for more than two hours after each concentration step. 
The two-dimensional dispersion of the sedimented particles was observed with optical microscopy (IX73 with an objective LUCPlanFLN 40X/0.60, Olympus) after the equilibration, exhibiting no further development of dispersed state such as bonding between particles or sticking of particles to the bottom surface, obtained by the analyses described below. 
The microscopy images were taken at a rate of 2~frames per second. 
A patch on a 2.7 $\mu$m particle is thin, and thus it is almost invisible in the images, see Fig.~\ref{sfig:i2}a.
Particle size and position were obtained from the binarized images, and the trajectory of each particle was determined from this data.

Figure~\ref{sfig:i2} shows a microscopy image and reconstructed images containing the information of the dispersed state obtained from 100 successive images. 
In the experiment, the particle area fraction in the two-dimensional plane was $\phi \simeq 0.6$, the number ratio of the patchy to non-patchy particle was 7:3, and the number of analyzed particles in an image was $\approx$~2400. 
In the reconstructed images we classified the dispersed state of a particle into 4 types.
When the variance of the distance between a particle and its neighbor is less than 0.1 $\mu$m$^2$, it is regarded as bonded (filled color circle with solid black lines denoting bonding). 
When the variance of the position of a particle is less than 0.03 $\mu$m$^2$, it is regarded as immobile. 
(Because of the noise in the images, the position of a particle and distance between aggregating particles obtained from the image processing fluctuate even when they were actually constant.)
The other particles are classified as monomers (filled gray circle), except for the particles that failed to be tracked (dotted open circle). 
A change in the so-defined type of a given particle was rarely observed under these experimental conditions: 
No change was found at 0.0 and 6.0 mM even in an hour observation, and one or two bond breaks were typically found at 4.0 mM in an hour, corresponding to $\sim$ 0.3 \% of all bonds. 

We found that a fairly large amount of patchy particles were bonded while being dispersed at 4.0 and 6.0 mM as shown in Figs.~\ref{sfig:i2}c and \ref{sfig:i2}d. 
In the figure most of the bonded particles are large, i.e. patchy, strongly suggesting that hydrophobic patches were bonded~\cite{chen2011directed}. 
The number fractions of bonded particles were 1.4, 39 and 36 \% for 0.0, 4.0 and 6.0 mM, respectively, corresponding to bonding of $\sim$ 60 \% of patchy particles at 4.0 and 6.0 mM. 
In addition, the weight fractions of dimers in all the bonded particles were 90 and 91\% for 4.0 and 6.0 mM, i.e., most of the bonded particles formed dimers. 
At these concentrations, the fraction of immobile particles were less than 1 \%. 

Further increase in salt concentration increased the fraction of bonded particles to some extent, that reached 48 and 63 \% at 8.0 and 10.0 mM, respectively,
although immobile particles increased more rapidly, 3.1 and 7.3 \% at the two concentrations. 
The weight fractions of dimers in all the bonded particles decreased, 75 and 63 \% at 8.0 and 10.0 mM, respectively, indicating that large aggregates increased more than dimers.
The bonding and immobilization of the small, non-patchy particles also increased in the aggregates and immobile particles at these concentrations, suggesting the increase of bonding between patch-silica, silica-silica and silica-glass surfaces. 
Note that the multi-bonding configurations allowed for the patch size would also contribute to the formation of the large aggregates.

In conclusion, this preliminary experiment demonstrates that colloidal particles with a hydrophobized patch can be bonded at any time in a dense dispersion by changing salt concentration, where $\sim$ 4.0 to 6.0 mM would be the best for achieving both bonding and dispersity of particles.

\end{document}